\documentclass[onecolumn, draftclsnofoot, 10pt]{IEEEtran}

\usepackage{amsmath}
\usepackage{amssymb}
\usepackage{array}
\usepackage{algorithmic}
\usepackage{algorithm}
\usepackage{theorem}
\usepackage{graphics}
\usepackage{epsfig}
\usepackage{subcaption}

\theorembodyfont{\rmfamily}
\theoremstyle{definition}\newtheorem{rem}{Remark}
\theoremstyle{definition}
\theoremstyle{definition}\newtheorem{cor}{Corollary}
\theoremstyle{definition}\newtheorem{defn}{Definition}
\theoremstyle{definition}\newtheorem{lem}{Lemma}
\theoremstyle{definition}\newtheorem{thm}{Theorem} 
\theoremstyle{definition}\newtheorem{prop}{Proposition} 
\theoremstyle{definition}
\theoremstyle{definition}

\title{Efficient Sensor Network Planning Method Using Approximate Potential Game}
\author{Su-Jin Lee, Young-Jin Park, and  Han-Lim Choi, ~\IEEEmembership{Member,~IEEE}}

\begin{document}
\maketitle
\begin{abstract}
	This paper addresses information-based sensing point selection from a set of possible sensing locations, which determines a set of measurement points maximizing the mutual information between the sensor measurements and the variables of interest. A potential game approach has been applied to addressing distributed implementation of decision making for cooperative sensor planning. When a sensor network involves a large number of sensing agents, the local utility function for a sensing agent is hard to compute, because the local utility function depends on the other agents’ decisions while each sensing agent is inherently faced with limitations in both its communication and computational capabilities. Accordingly, a local utility function for each agent should be approximated to accommodate limitations in information gathering and processing. We propose an approximation method for a local utility function using only a portion of the decisions of other agents. The part of the decisions that each agent considers is called the neighboring set for the agent. The error induced by the approximation is also analyzed, and to keep the error small we propose a neighbor selection algorithm that chooses the neighbor set for each agent in a greedy way. The selection algorithm is based on the correlation information between one agent's measurement selection and the other agents' selections. Futhermore, we show that a game with an approximate local utility function has an $\epsilon$-equilibrium and the set of the equilibria include the Nash equilibrium of the original potential game. We demonstrate the validity of our  approximation method through two numerical examples on simplified weather forecasting and multi-target tracking.
  
\end{abstract}

%===============================================================
\section{Introduction}
\vspace{10pt}
The goal of cooperative sensor network planning problems is to select the sensing locations for a sensor network so that the measurement variables taken at those locations give the maximum information about the variables of interest. This problem can be formulated as an optimization problem with the global objective function of mutual information between the measurement variables and the variables of interest \cite{Choi2015_IEEETCST}, \cite{Gopalakrishnan2011_ACM}, \cite{Hoffmann2010_IEEETAC}, \cite{Krause2008_JMLR}. For the distributed/decentralized implementation of the optimization problem, there are two main research directions.% aimed at designing the procedure finding out the most informative sensing locations. 
The two directions can be differentiated by the number of steps required to solve a local optimization problem for each agent until a solution is obtained. One direction can be described as a single run algorithm, which includes local greedy and sequential greedy decisions \cite{Krause2008_JMLR}, \cite{Nguyen2016_IEEETCST}. While these algorithms are simple to implement, and especially a sequential greedy algorithm guarantees the worst-case performance when the objective function satisfies submodularity. However, they are subject to some limitations. Since each agent selects the sensing locations by solving only one problem, these single run algorithms do not fully take advantage of possible information flows, and thus the decisions can be arbitrarily suboptimal. The other direction is an iterative algorithm which generates a sequence of solutions to converge to an approximate optimal solution \cite{Choi2015_IEEETCST}, \cite{Hoffmann2010_IEEETAC}, \cite{Grocholsky2002_Thesis}. An iterative method solves the optimization problem approximately at first, and then more accurately with an updated set of information as the iterations progress \cite{Boyd2011_FTML}. A game-theoretic method is one of the iterative algorithms, which finds a solution through a decision making process called a repeated game, i.e., the same set of games being played until converging to a solution. Especially, a potential game approach provides a systematic framework for designing distributed implementation of multiagent systems and many learning algorithms that guarantees convergence to an optimal solution \cite{Marden2009_IEEETAC, Marden2007_ICAAMS}. 

In \cite{Choi2015_IEEETCST}, we adopted a potential-game approach to a sensor network planning. Potential games have been applied to many multiagent systems for distributed implementation due to their desirable static properties (e.g., existence of a pure strategy Nash equilibrium) and dynamic properties (e.g., convergence to a Nash equilibrium with simple learning algorithms) \cite{Candogan2011_MathOR}, \cite{Marden2009_IEEETAC}. The formulation of a multiagent problem as a potential game consists of two steps: (1) game design in which the agents as selfish entities and their possible actions are defined; and (2) learning design which involves specifying a distributed learning algorithm that leads to a desirable collective behavior of the system \cite{Gopalakrishnan2011_ACM}. 
For game design, we proposed the conditional mutual information of the measurement variables conditioned on the other agents' decisions as a local utility function for each agent. This conditional mutual information is shown to be aligned with the global objective function for a sensor network, which is mutual information between the whole sensor selection and the variables of interest. For a learning algorithm, joint strategy fictitious play is adopted.  With these two design steps we showed that the potential game approach for distributed cooperative sensing provides better performance than other distributed/decentralized decision making algorithms, such as the local greedy and the sequential greedy algorithms.

However, the computation of the local utility function for each agent requires extensive resources. This computational burden results in part from the complexity of the local utility function itself, and in part from the dependency of the function on all the agents' decisions. This decision sharing among all the sensors also cause communication overload on a network. To address this issue, an approximate local utility that only depends on the neighboring agents' decisions was suggested and investigated \cite{Choi2015_IEEETCST}. Since it is not possible for each agent to know the decisions from the other agents’ actions for a sensor network consisting of a large number of sensors, this approximate local utility function also enables us to improve communication efficiency. Thus, the local utility function for each agent is computed by considering only the actions of its neighboring agents. Here, the neighboring agents for each agent are defined by a set of agents that are located within a prespecified distance from the agent. When selecting neighboring agents for each agent, the correlation between the measurement variables is not considered at all. However, in some cases, such as a weather forecast example the measurements taken at close locations can have little correlation with each other. Thus, the optimality gap of the potential game with the neighboring agents can become larger than the potential game with the full information about the other agents and even the sequential greedy algorithm, because the error incurred by neglecting the correlated variables is not negligible. 

This work presents an approximation method for computing local utility function to address this computation and communication issues. We first propose a simple neighbor selection algorithm based on network topology, which was included in our previous work \cite{Choi2013_ACC}. In addition to this method, we also propose a greedy neighbor selection algorithm to consider the correlation structure of the information space in which the cooperative sensing decision is made. The greedy selection algorithm for each agent adds a neighboring agent one by one, which has maximum mutual information about the variables of interest conditioned on the measurement variables of the agent and its pre-selected neighbors. With the determined neighbor set for each agent, we propose a similar approximation method for computing a local utility function to the previous work. Two numerical examples are presented to show the feasibility and validity of our approximation algorithm. The first example on multi-target tracking demonstrates the usefulness of a simple selection algorithm, since the correlation between sensor measurements depends on the distance between those sensing points. A numerical example on idealized weather forecasting is presented, showing that the approximation utility functions depending on the neighbor set selected by the proposed greedy algorithm outperform the approximation method with neighboring agents which are located close in Euclidean distance. 

In Section \ref{sec:MIformulation}, we give an overview of the sensor network planning problem. In Section \ref{sec:PGformulation}, we present a background for a game-theoretic approach and the problem formulation using a potential game. We desrcibe our approximation algorithm to address computational and communicational issues of the approach in Section \ref{sec:MainResults}. Simulated examples are presented in Section \ref{sec:Simulations}.

\section{Sensor Network Planning Problems} %Problem Formulation
\label{sec:MIformulation}
This section describes a sensor network planning problem. A sensor network model is defined, and provides a formulation as an optimization problem with a global objective function using mutual information. 

\subsection{Sensor Network Model}
\begin{figure}[t]
	\centering
	\includegraphics[width=.9\columnwidth]{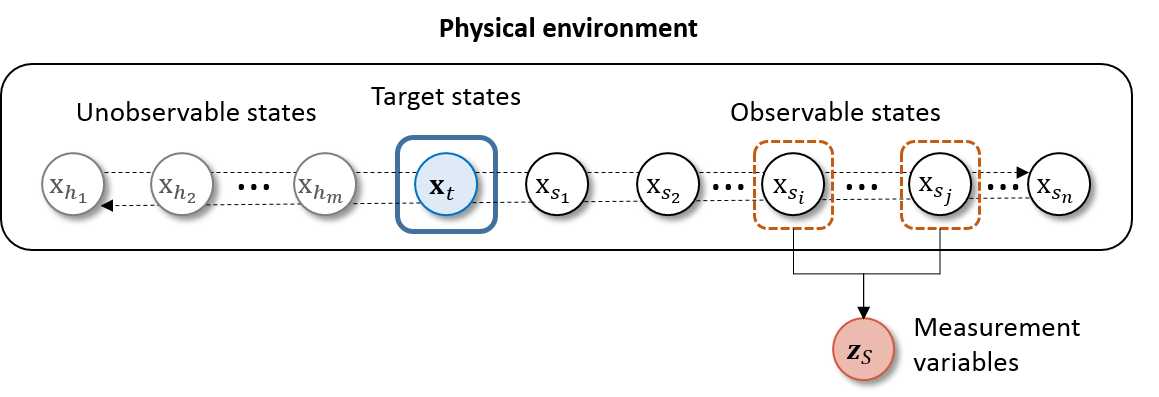}
	\caption{A graphical model for representing probabilistic relationships between states and measurement variables}
	\label{fig:relation_of_states}
\end{figure}

%[verification variables]
The primary task of a sensor network is to collect information from a physical environment in order to estimate the variables of interest, called \textit{target states}. %(verification variables ??). 
The target states $\mathbf{x}_t$  are the set of variables of interest we want to know through the measurements. They are a part of the state variables representing the physical environment (see Fig. \ref{fig:relation_of_states}). In a problem of our concern, the physical environment is represented with a finite set of measurable states, $\mathbf{x}_{\mathcal{S}}=[\mathrm{x}_{s_1 },\mathrm{x}_{s_2},~\dots,~ \mathrm{x}_{s_M}]$ and the target states $\mathbf{x}_t$ in spatial-temporal space. The subscript $\mathcal{S}=\{s_1, s_2,~\dots,~s_M\}$ denotes a set of possible sensing locations for a sensor network, and is referred to as a search space, and each element $\mathrm{x}_s$  represents the state variable at the sensing location $s\in \mathcal{S}$. 

%[sensor network & measurement model]
We consider a sensor network consisting of $N$ sensing agents deployed in the search space.  A sensing agent can be a vehicle carrying sensors on it, such as an unmanned aerial vehicle(UAV). In the case of sensor nodes installed at fixed locations, a sensing agent can be considered as a virtual agent that selects a set of sensor nodes to turn on.  Due to the dynamic constraints on the sensing agents, the search space for each sensing agent is a subset of the whole search space $\mathcal{S}_i\subset\mathcal{S}$. %\footnote{In case of sensor nodes installed at the fixed locations, the search space for a virtual agent can be the whole search space. If the search space is divided and a predefined number of sensors should be turned on in each  }.  
For a mobile sensor network, the search space for each sensing agent can be obtained by discretizing the search space which is available at the next time step.  

The sensor measurements for the $i$-th agent at the sensing point $s_i \in \mathcal{S}_i$ is a function of the state variable at the point and the variables of interest:
\begin{equation} \label{eq:measuremodel}
\mathrm{z}_s = h(\mathrm{x}_s, \mathbf{x}_t) + w_s, ~ w_s\sim \mathcal{N}(0,R_s)
\end{equation}
where $w_s$ is white noise uncorrelated with any other random variables. The observation function $h(\cdot)$ could be a nonlinear or linear mapping of the states and the target states. In a target tracking problem, the variables of interest are the location of the target, and if sensors  observe the bearing or range to the target, then the observation function $h(\cdot)$ is expressed as a function of relative geometry of s and the target location $\mathbf{x}_t$. In a weather forecasting problem, $h(\cdot)$ depends only on the state variable itself at the sensing location $s$, such as temperature and pressure, then the observation model becomes $\mathrm{z}_s = \mathrm{x}_s + w_s$. 

%[verification variables & measurement variable]
Since the target states $\mathbf{x}_t$ are correlated with the states in the search space and the sensors make observations of the states in the search space corrupted with some noise, sensor measurements are correlated with the target states and have information about the  target [\ref{eq:measuremodel}]. The correlation between the target states and the other states in a search space can be expressed as a joint probability distribution $p(\mathbf{x}_t, \mathbf{x}_{\mathcal{S}})$ of two sets of random variables, $\mathbf{x}_t$ and $\mathbf{x}_{\mathcal{S}}$. Accordingly, the correlation between the target states and the measurements can be represented by the joint probability distribution $p(\mathbf{x}_t, \mathbf{z}_{\mathcal{S}})$. To obtain the probability distribution of whole states, the states are represented with a probabilistic model, and moreover if the states evolve with time, we also should establish a dynamic model of the states. With the models related to the states, a measurement model will give an expression for the joint probability distribution of the target states and the measurement. 

\subsection{Sensor Planning Problem for Maximum Information} 
%[problem formulation as optimization]
% -- mutual information for quantification of uncertainty)
The goal of sensor network planning problems is to find out the optimal sensing locations that maximize the information reward about the target states. Here, the information reward can be thought of as a reduction in the uncertainty of the variable of interest due to the information included in the measurement variables. This uncertainty reduction is quantified with mutual information, which is the difference between the prior and posterior entropy of $\mathbf{x}_t$ conditioned on $\mathbf{z}_{s}$ \cite{Cover1991_Book}. 

\begin{equation}  \label{eq:mutinf}
  \mathcal{I}(\mathbf{x}_t; \mathbf{z}_{s}) = \mathcal{H}(\mathbf{x}_t)-\mathcal{H}(\mathbf{x}_t|\mathbf{z}_s)
\end{equation}
where 
\begin{eqnarray}  \label{eq:entropy}
  \mathcal{H}(\mathbf{x}_t)&=& -\int p(\mathbf{x}_t) \log p(\mathbf{x}_t) d\mathbf{x}_t \\
  \mathcal{H}(\mathbf{x}_t|\mathbf{z}_s) &=& -\int p(\mathbf{x}_t, \mathbf{z}_s) \log p(\mathbf{x}_t|\mathbf{z}_s) d\mathbf{x}_t d\mathbf{z}_s
\end{eqnarray}
$\mathcal{H}(\mathbf{x}_t)$ is the entropy of the target states distribution, $\mathcal{H}(\mathbf{x}_t|\mathbf{z}_s)$ is the conditional entropy of the distribution when conditioning on the measurements, and $\mathcal{I}(\mathbf{x}_t; \mathbf{z}_s)$ is the mutual information between the target states and the measurements. The entropy of a random variable is a metric of its uncertainty. As shown in the above definition, the entropy is not dependent on the value of the random variable itself but rather depends on the shape of the distribution. In the case of a Gaussian distribution, the entropy is computed by the determinant of the covariance matrix $P(\mathbf{x}_t)$, not using its mean value.  
\begin{equation} \label{eq:entropy:normal}
  \mathcal{H}(\mathbf{x}_t)=\frac{n_t}{2}\log(2\pi e)+\frac{1}{2}\log(|P(\mathbf{x}_t)|)
\end{equation}
where $n_t$ is the dimension of the random vector $\mathbf{x}_t$. When the states are jointly Gaussian and the measurement model is linear and the observation noise is Gaussian, then the mutual information can be expressed with the prior and posterior covariance matrices
\begin{equation} \label{eq:mutinf:normal}
  \mathcal{I}(\mathbf{x}_t; \mathbf{z}_{s})=\frac{1}{2}\log(|P(\mathbf{x}_t)|)-\frac{1}{2}\log(|P(\mathbf{x}_t|\mathbf{z}_s)|)
\end{equation}

% -- optimization using mutual information 
Since a sensor network has limited resources, the system should select the set of sensing points that give the maximum information about the target states to reduce the number of observations. Therefore, a sensor network planning problem can be stated as selecting the most informative set of sensing points $\mathbf{s}_{1:N}=\{s_1, s_2, \dots, s_N\}$ over the search space $\mathcal{S}$. 
\begin{equation}
  \label{eq:globalopt}
   \mathbf{s}^*_{1:N} = \arg \max_{\mathbf{s}_{1:N}:s_i\in\mathcal{S}_i} \mathcal{I}(\mathbf{x}_t;\mathbf{z}_{\mathbf{s}_{1:N}})
\end{equation}
Note that $i$-th sensing location $s_i$ is selected from its designated region $\mathcal{S}_i$. $\mathbf{z}_{\mathbf{s}_{1:N}}=[\mathrm{z}_{s_1}, \dots, \mathrm{z}_{s_N}]$ is a random vector that represents the measurement variables taken at the locations $\mathbf{s}_{1:N} \subset \mathcal{S}$. 
In a centralized approach, the problem is a combinatorial optimization: it computes the mutual information for every combination of sensing points and then chooses the set giving the maximum mutual information. Thus, the computational burden increases exponentially with the number of sensing agents and the size of the search space. To address this combinatorial computational complexity, we proposed a potential game-based approach and showed the validity of the proposed method through numerical examples. In the following section, we will review a game-theoretic framework for cooperative sensor planning and analyze the complexity of this formulation. 

\section{Sensor Network Planning as a Potential Game} % Problem Formulation
\label{sec:PGformulation}
Potential games are applied to many engineering optimization problems (such as cooperative control \cite{Marden2009_IEEETSMC} and resource allocation \cite{Choi2016_DARS}) due to their  static (existence of Nash equilibrium) and dynamic properties (simple learning algorithm) \cite{Candogan2011_MathOR}. This section provides the required game-theoretic background used to develop the results in this paper and describes a potential game formulation of a sensor network planning problem. Lastly, we present limitations in applying the approach to a large sensor network. 
%This section describes a game-theoretic framework for solving the sensor network planning problem. First, a sensor network model is defined, and provides a formulation as an optimization problem with a global objective function using mutual information. Lastly, a game theoretic framework is presented for formulating the optimal problem as a potential game. 

%______________________________________
\subsection{Game-Theoretic Architecture}
Consider a finite game in strategic form consisting of three components \cite{Fudenberg1991_Book}. 
\begin{itemize}
\item A finite set of players(agents): $\mathcal{N}=\{1,2,\dots,N\}$.
\item Strategy spaces: a finite set of actions(strategies) $\mathcal{S}_i$, for each player $i\in\mathcal{N}$.
\item Utility functions: a utility (payoff) function $U_i:\mathcal{S}\rightarrow\mathbb{R}$, for each player  $i\in\mathcal{N}$.
\end{itemize}
Accordingly, a finite strategic form game instance is represented by the tuple $\langle\mathcal{N},\allowbreak\{\mathcal{S}_i\}_{i\in\mathcal{N}},\allowbreak\{U_i\}_{i\in\mathcal{N}}\rangle$. In this setting, each player has an action set that the player can select, and has a preference structure over the actions according to its utility function $U_i$. A joint strategy space of a game is denoted by $\mathcal{S}=\prod_{i\in\mathcal{N}}\mathcal{S}_i$, which represents a set of all possible combinations of actions for all players to choose at a time. $\mathbf{s}=(s_1,s_2,\dots,s_N)\in\mathcal{S}$ is the collection of strategies of all players, called a strategy profile, where $s_i\in\mathcal{S}_i$ denotes the strategy chosen by player $i\in\mathcal{N}$. For notational convenience, $\mathbf{s}_{-i}=(s_1,\dots,s_{i-1},s_{i+1},\dots,s_N)$ denotes the collection of actions of players other than player $i$. With this notation, a strategy profile is expressed as $\mathbf{s}=(s_i,\mathbf{s}_{-i})$. 

A utility function $U_i(s_i, \mathbf{s}_{-i})$ for player $i$ reflects the preference of player $i$ over its possible actions $\mathcal{S}_i$. Given other players' actions, each player would prefer an action that gives the maximum payoff. 
If every player selects the action with the maximum payoff given other players' actions and the chosen action is consistent with the beliefs that other players assumed about the player's action, and it is also true for all the players, then no player would change his action in this strategy profile. 
This follows a solution concept in a non-cooperative game, a Nash equilibrium. Formally, a strategy profile $\mathbf{s}^*\in\mathcal{S}$ is a pure Nash equilibrium if 
\begin{equation}
  \label{eq:def:NE}
  U_i(s_i,\mathbf{s}_{-i}^*)- U_i(s_i^*,\mathbf{s}_{-i}^*) \leq 0
\end{equation}
for every $s_i\in\mathcal{S}_i$ and every player $i\in\mathcal{N}$.
A  Nash equilibrium is a strategy profile in which no player can improve its payoff by deviating unilaterally from its profile. 

\iffalse
To address games that are close in terms of payoffs to potential games, we use $\epsilon$-equilibrium solution concept. A strategy profile $\mathbf{s}^{\epsilon}\in\mathcal{S}$ is an $\epsilon$-equilibriu if 
\begin{equation}
  \label{eq:def:nearNE}
  U_i(s_i,\mathbf{s}_{-i}^{\epsilon})-  U_i(s_i^{\epsilon},\mathbf{s}_{-i}^{\epsilon}) \leq \epsilon
\end{equation}
for every $s_i\in\mathcal{S}_i$ and every player $i\in\mathcal{N}$. 
\fi 

%\paragraph{Potential Game}
A potential game is a non-cooperative game in which the incentive of the players changing their actions can be expressed by a single function, called the potential function. That is, that the player tries to maximize its utility is equivalent to maximizing the global objective for a set of all the players \cite{Monderer1996_Games}.
\begin{defn}
  \label{def:PG}
  A finite non-cooperative game $\mathcal{G}=\langle \mathcal{N}, \{\mathcal{S}_i\}_{i\in\mathcal{N}}, \{U_i\}_{i\in\mathcal{N}}  \rangle$ is a potential game if there exists a scalar function $\phi:\mathcal{S}\rightarrow\mathbb{R}$ such that \\
  \begin{equation}
    \label{eq:def:PG}
    U_i(s_i',\mathbf{s}_{-i})-U_i(s_i'',\mathbf{s}_{-i})=\phi(s_i',\mathbf{s}_{-i})-\phi(s_i'',\mathbf{s}_{-i})
  \end{equation}
  for every $i\in\mathcal{N}$, $s_i',s_i''\in\mathcal{S}_i$, $\mathbf{s}_{-i}\in\mathcal{S}_{-i}$. The function $\phi$ is referred to as a potential function of the game $\mathcal{G}$.
\end{defn}
The property of a potential game in (\ref{eq:def:PG}) is called a perfect alignment between the potential function and the player's local utility functions. In other words, if a player changes its action unilaterally, the amount of change in its utility is equal to the change in the potential function.
 
Potential games have two important properties \cite{Candogan2011_MathOR}. The first one is that the existence of pure strategy Nash equilibrium is guaranteed. Since in a potential game the joint strategy space is finite, there always exists at least one maximum value of the potential function. This strategy profile maximizing the potential function locally or globally is a pure Nash equilibrium. Hence, every potential game possesses at least one pure Nash equilibrium. 
The second important property is the presence of well-established learning algorithms to find a Nash equilibrium by repeating a game. Many learning algorithms for potential games are proven to have guaranteed asymptotic convergence to a Nash equilibrium \cite{Marden2009_IEEETSMC}. 

In this paper, our main focus is to approximate utility functions of a potential game. To address strategy profiles in a near potential game with approximate local utility functions, a near Nash equilibrium is introduced. A strategy profile $\mathbf{s}^\epsilon$ is an $\epsilon$-equilibrium if 
\begin{equation}
  \label{eq:def:nearNE}
  U_i(s_i,\mathbf{s}_{-i}^\epsilon) - U_i(s_i^\epsilon,\mathbf{s}_{-i}^\epsilon) \leq \epsilon
\end{equation}
for every $s_i\in\mathcal{S}_i$ and every player $i\in\mathcal{N}$. In a Nash equilibrium, every player chooses the action that gives the payoff equal to or greater than the payoffs for choosing other actions.
 In a near Nash equilibrium, there can be an action which gives a better payoff by no more than $\epsilon$. When $\epsilon$ is equal to zero, an $\epsilon$-equilibrium is a Nash equilibrium. 

\iffalse
 Two games that are close each other in their utilities have similar equilibrium. The following lemma shows the extent of how close the $\epsilon$-equilibria of two games are \cite{Candogan2011_MathOR}.
 
\begin{lem} 
  \label{lem:nearNE}
   Consider two games $\mathcal{G}=\langle\mathcal{N},\{S_i\}_{i\in\mathcal{N}},\{U_i\}_{i\in\mathcal{N}}\rangle$ and $\tilde{\mathcal{G}}=\langle\mathcal{N},\allowbreak\{\mathcal{S}_i\}_{i\in\mathcal{N}},\allowbreak\{\tilde{U_i}\}_{i\in\mathcal{N}}\rangle$ which differ only in their utility functions. If $|U_i(\mathbf{s})-\tilde{U_i}(\mathbf{s})|\leq\Delta_u$ for every $i\in\mathcal{N}$ and $\mathbf{s}\in\mathcal{S}$. Then, every $\tilde{\epsilon}$-equilibrium of $\tilde{\mathcal{G}}$ is an $\epsilon$-equilibrium of $\mathcal{G}$ for some $\epsilon \leq 2\Delta_u + \tilde{\epsilon}$ (and vice versa).
\end{lem} 
Specifically, if $\mathcal{G}$ has a Nash equilibrium, then the other game has an $\epsilon$-equilibrium, such that $\epsilon\leq 2\Delta_u$. Therefore, if we make a game sufficiently close to the existing game with a Nash equilibrium, then an equilibrium property of an approximate game can be identified. 
\fi

\subsection{Cooperative Sensing as a Potential Game}
From a game-theoretic perspective, each sensing agent is considered a player in a game who tries to maximize its own local utility function, $
U_i (s_i, \mathbf{s}_{-i})$, where $s_i$ is the set of sensing locations for sensor $i$, and $\mathbf{s}_{-i}$ represents the set of sensing locations other than sensor $i$'s selections. In our previous work \cite{Choi2015_IEEETCST}, we showed that the conditional mutual information of sensor $i$'s measurements conditioned on the other agents’ sensing decisions leads to a potential game with a global objective function
 \begin{equation}
     \label{eq:pg:wlu:potential} \phi({s_i},{\mathbf{s}_{-i}})=\mathcal{I}(\mathbf{x}_t;\mathrm{z}_{s_i},\mathbf{z}_{\mathbf{s}_{-i}})=\mathcal{I}(\mathbf{x}_t;\mathbf{z}_{\mathbf{s}_{1:N}}).
  \end{equation}
The local utility function can be represented by
\begin{equation}
    \label{eq:pg:wlu:local}
    U_i({s_i},{\mathbf{s}_{-i}})=\mathcal{I}(\mathbf{x}_t;\mathrm{z}_{s_i}|\mathbf{z}_{\mathbf{s}_{-i}}).
  \end{equation}
With this local utility function, the designed potential game is solved by repeating the game for $t\in\{0, 1, 2, \dots \}$. At each time step $t$, each agent $i\in\mathcal{N}$ chooses an action according to a specific learning rule (such as fictitious play, better/best response and log-linear learning \cite{Fudenberg1998_Book}), which is generally represented as a probability distribution $p_i(t)$ over $i$'s action set $\mathcal{S}_i$. $p_i(t)$ is obtained from the utility values of agent $i$  based on the other agents' decisions up to the previous stages. The general structure of learning algorithms is summarized in Algorithm \ref{alg:learning_game}. %The learning algorithms are varied according to a specific update rule $F_i$.%$F_i$ represents a specific updated rule. 

\begin{algorithm}[b] 
  \caption{{\sc Learning Algorithm} ($U_i, F_i$)}
  \begin{algorithmic} \label{alg:learning_game}
   \STATE Choose an initial action with some specific rule. 
   \WHILE{ Convergence criteria not satisfied}
    \FOR{ $i \in \{1,\dots, N\}$}
    \STATE Update the strategy according to the learning rule, $p_i(t)=F_i(\mathbf{s}(0),\dots,\mathbf{s}(t-1);U_i)$
    \STATE Choose an action according to $p_i(t)$ %Perform local optimization at each agent $i$
   \ENDFOR
   \ENDWHILE
\end{algorithmic}
\end{algorithm}

The learning algorithms are varied according to a specific update rule $F_i$. % \cite{Fudenberg1998_Book}. (such as fictitious play, better/best response and log-linear learning \cite{Fudenberg1998_Book}). 
To solve a sensor network planning problem, we adopted joint strategy fictitious play (JSFP) as a learning rule \cite{Marden2009_IEEETAC}.
In JSFP, each agent assumes that other agents play randomly according to the joint empirical frequencies, $f_{-i}(\mathbf{s}_{-i};t)$, which represents the frequency with which all players but $i$ have selected a joint action profile $\mathbf{s}_{-i}$ up to stage $t-1$. In a strategy update step at each stage $t$, a player computes the expected local utility for action $s_i \in \mathcal{S}_i$ based on the empirical frequency distribution of its opponents as
\begin{equation*}
U_i(s_i; t) = \mathbb{E}_{f_{-i}(t)} \left[ U_i (s_i, \mathbf{s}_{-i}) \right]
\end{equation*}
In \cite{Marden2009_IEEETAC}, it is shown that the expected utilities $U_i(s_i; t)$ for each  $s_i \in \mathcal{S}_i$ can be expressed with a simple recursion rule 
\begin{eqnarray} \label{eq:learning:jsfp}
  U_i (s_i; t) &=& \frac{1}{t} \sum_{\tau = 0}^{t-1} U_i (s_i, \mathbf{s}_{-i}(\tau)) \notag\\
  &=& \frac{t-1}{t} U_i (s_i; t-1) + \frac{1}{t} U_i (s_i, \mathbf{s}_{-i}(t-1)).
\end{eqnarray}
Having computed $U_i(s_i;t)$ for all $s_i\in\mathcal{S}_i$, agent $i$ selects the best action that gives the maximum expected utility. Although JSFP does not guarantee convergence to a pure Nash equilibrium even for a potential game, a JSFP with inertia was proven to reach a Nash equilibrium. 
In our previous work \cite{Choi2015_IEEETCST}, we showed that the JSFP for a sensor network planning problem can be converged to a Nash equilibrium by using \cite[Theorem 2.1]{Marden2009_IEEETAC} and specifying the termination condition of the algorithm.

%===============================================================
\subsection{Computational Complexity Analysis} \label{sec:comp_time_analysis}
%In this section, we focus on the viability of a potential game approach to sensor network planning by analyzing the computation time to find a Nash equilibrium of the potential game. 
% 센서 네트워크 planning problem을 포텐셜 게임으로 formulation - conditional mutual information으로 정의 
% repeated game을 통해서 포텐셜 게임의 해, NE를 구함. 
% 포텐셜 게임의 해인 내쉬 평형점은 반복게임을 통해서 구해진다. 
% 반복 게임의 정의 
% 따라서 계산량 = 반복 게임 play 수 X 가능한 action 개수 X utility 한 번 계산량 
% 포텐셜 게임으로 formulation해서 exhaustive search 보다 적은 계산량으로 해를 구할 수 있음을 보였다. 하지만 여전히 계산량에 대한 이슈가 있다. 
As described in \cite{Choi2011_IEEETCST}, the potential game-based approach to cooperative sensor planning enables one to find out the informative set of sensing locations with much less computation than an exhaustive search. However, a conditional mutual information conditioned on the other agents' sensing locations has limitations when applied to a large sensor network, because every agent needs information about the decisions made by the other agents in the previous stage. If the communication network is strongly connected, one agent's decision can be forwarded to all the other agents, and then the next stage can be initiated after this information exchange has happened. This decision sharing among all the agents incus substantial communication cost/traffic. In addition to this communicational burden on a network, the conditional mutual information itself increases computation load on each agent's processing unit for obtaining utility values. 
In this section, we will analyze the computation complexity for JSFP algorithm  applied to sensor network planning in more detail.  

As shown in Algorithm \ref{alg:learning_game}, the distributed solution of a potential game is obtained by repeating a  one-shot game until it converges to equilibrium. In JSFP, at each stage $t$, agent $i\in \mathcal{N}$ updates its expected utility values $U_i(s_i; t)$ over the possible actions $s_i\in\mathcal{S}_i$, and then chooses the best action that gives the maximum expected utility. To obtain $U_i(s_i; t)$, an agent should first compute the utility values $U_i(s_i, \mathbf{s}_{-i}(t-1))$ based on the previous decisions of the other agents as shown in (\ref{eq:learning:jsfp}), and then update the expected utility with the weighted sum of the previous average utility $U_i(s_i; t-1)$ and the utility $U_i(s_i, \mathbf{s}_{-i}(t-1))$ obtained from a one-step look-back scheme. The resulting computation time for agent $i$ to reach a solution is then
%In order to obtain the preferenece $p_i(t)$ over actions, each agent computes its local utility function for every possible actions with the information up to the present time{\footnote{In some learning algorithms, every agent doesn't calculate its local utility function over possible actions. For example, the learning rules with inertia allow an agent to keep the previous action unchanged with some probability, and the agent is not required to compute its utility at that step. In our work, we consider a worst-case computation.}}. The total computation time for agent $i$ to reach a solution is then  
%A learning rule would find a Nash equilibrium to find a Nash equilibrium would calculate the preference $p_i(t)$ followed by choosing an action according to a comparison between utility values until convergence to the equilibrium. The total computation time for a learning process of agent $i$ is then  
\begin{equation} \label{eq:learning:time}
\hat{T}_i = L  N_{\mathcal{S}_i} N_{p_i} T_{U_i} 
\end{equation}
where $L$ is the number of repeating a game until convergence, $N_{\mathcal{S}_i} $ is the size of an action set $\mathcal{S}_i$ and $N_{p_i}$ is the number of computation of the local utility for each play of a game. $T_{U_i} $ corresponds to the time taken to calculate the local utility function for a given strategy profile. Here, $N_{p_i}$ is constant for the specific learning algorithm and $L$ is related to the convergence rates of the learning algorithms for a potential game. Although its relation to the game size has not been fully established for a general potential game, it has been shown that many of the learning rules converge to a solution within reasonable time for specific potential game structures \cite{Borowski2013_IEEECDC} and especially joint strategy fictitious play applied to a sensor network planning problem has shown to converge to a  close-to-optimal solution with computation cost less than greedy strategies \cite{Choi2015_IEEETCST}. Thus, we focus on the term $T_{U_i}$ to analyze the computation time for a potential game approach. 

The main reason for the computation load of $T_{U_i}$ is due to conditioning the decisions from all of the agents. For example, in case of a multivariate normal distribution, a local utility function (\ref{eq:pg:wlu:local}) can be rewritten using the backward scheme in \cite{Choi2011_IEEETCST} and the mutual information for Gaussian variables (\ref{eq:mutinf:normal}) with additive white Gaussian measurement noise,
\begin{equation} \label{eq:pg:wlu:localbw}
\begin{split}
U_i(s_i,{\mathbf{s}_{-i}})&=\mathcal{I}(\mathbf{x}_t;\mathrm{z}_{s_i}|\mathbf{z}_{\mathbf{s}_{-i}})
=\mathcal{H}(\mathrm{z}_{s_i}|\mathbf{z}_{\mathbf{s}_{-i}}) - \mathcal{H}(\mathrm{z}_{s_i}|\mathbf{x}_t,\mathbf{z}_{\mathbf{s}_{-i}}) \\
&=\frac{1}{2} \log | P(\mathrm{z}_{s_i}|\mathbf{z}_{\mathbf{s}_{-i}})| - \frac{1}{2}\log | P(\mathrm{z}_{s_i}|\mathbf{x}_t, \mathbf{z}_{\mathbf{s}_{-i}})| \\
&= \frac{1}{2}\log  \left| P(\mathrm{z}_{s_i}) - P(\mathrm{x}_{s_i},\mathbf{x}_{\mathbf{s}_{-i}})P(\mathbf{z}_{\mathbf{s}_{-i}})^{-1}P(\mathbf{x}_{\mathbf{s}_{-i}},\mathrm{x}_{s_{i}}) \right|  \\
&\quad - \frac{1}{2}\log \left|P(\mathrm{z}_{s_i}|\mathbf{x}_t) -P(\mathrm{x}_{s_i},\mathbf{x}_{\mathbf{s}_{-i}}|\mathbf{x}_t)P(\mathbf{z}_{\mathbf{s}_{-i}}|\mathbf{x}_t)^{-1}P(\mathbf{x}_{\mathbf{s}_{-i}},\mathrm{x}_{s_{i}}|\mathbf{x}_t) \right|
\end{split}
\end{equation}
where $P(\mathbf{z}_{\mathbf{s}})= P(\mathbf{x}_{\mathbf{s}}) + R_{\mathbf{s}}$, $P(\mathbf{z}_{\mathbf{s}}|\mathbf{x}_t)=P(\mathbf{x}_{\mathbf{s}}|\mathbf{x}_t) + R_{\mathbf{s}}$ denote the covariance matrices of measurement variables at sensing selections $\mathbf{s}$, and $R_{\mathbf{s}}$ denotes the measurement noise covariance. As shown above, the computation of the inverse of matrices relating to the other agents' decisions is needed to obtain a local utility function. Inversion of $n \times n$ symmetric positive matrix requires approximately $\frac{2}{3}n^3$ floating-point operations \cite{Andersen2002_LectureNotes}, and thus computation time for the conditional mutual information increases proportional to the cubic of the number of sensing agents $(\mathcal{O}(N^3))$. For a large sensor network, the computation of a utility function for an agent becomes intractable. Therefore, to keep the game approach scalable as the number of sensors increases, we need to approximate the local utility function. 

%However, the computation of the local utility functions requires a lot of resources. This computational burden results in part from the complexity of the local utility function itself, and in part from the dependency of the function on all of the agents’ decisions.  In a sensor network involving a large number of sensors, each sensing node has limited capability in communication, thus it is not possible for each agent to know the decisions from the other agents’. As a result, we need to approximate the local utility function for the sensing agent. 

%\paragraph{Specialization to Joint Strategy Ficitious Play \cite{Marden2009_IEEETAC}}
%To be more specific, we consider joint strategy fictitious play 

%\paragraph{Gaussian Case}
% Gaussian case - backward 
% \cite{Choi2011_IEEETCST}와 같은 방법으로 각 게임에서 결정하기 위해 필요한 계산량 계산 

%\paragraph{Non-Gaussian Case}
% Non-gaussian case - particle filter 
% cite{Hoffmann2010_IEEETAC}

%===============================================================
\section{Approximate Local Utility Design}
\label{sec:MainResults}
%The computaional burden 
In this section, an approximation method is proposed to reduce the computational and communicational burden of computing the local utility function. The method modifies the form of a local utility function itself by removing some of the conditioning variables, which represent the decisions of other agents.  
%______________________________________
\subsection{Approximate Utility Function} % Change the title: this doesn't describe the algorithm 
The proposed approximation method for computing the local utility function is to reduce a number of the conditioning variables using the correlation structure of the information space, making the local utility function depend on the part of the decisions. 
\begin{equation}
\label{eq:nearpg:wlu:local}
\tilde{U}_i (s_i, \mathbf{s}_{N_i})=\mathcal{I}(\mathbf{x}_t;\mathrm{z}_{s_i}|\mathbf{z}_{\mathbf{s}_{N_i}})
\end{equation}
where ${\mathbf{s}_{N_i}}$ is the set of the measurement selections correlated with sensing agent $i$'s decision, referred to as a neighbor set of sensing agent $i$ and $\mathbf{z}_{\mathbf{s}_{N_i}}$ denotes the corresponding measurement variables. The neighbor set ${\mathbf{s}_{N_i}}$ is a subset of ${\mathbf{s}_{-i}}$. To reduce the computation, the neighbor set should be a strict subset of ${\mathbf{s}_{-i}}$. In some applications, this allows for less memory requirement for each agent (as, for example, an agent only needs to carry covariance matrix for the random vector associated with its own and neighboring agents' sensing regions). The following lemma shows the quantification of the error incurred by the approximation. 

\begin{lem} \label{lem:nearpg:error}
Let $\Delta_{U_i}$ denote the difference between the approximate local utility for sensor $i$ and the true value of (\ref{eq:pg:wlu:local}), then 
\begin{subequations}
\label{eq:nearpg:wlu:error}
\begin{align}
  \Delta_{U_i} &= c_1(\mathbf{s}_{-i}) - \mathcal{I}(\mathbf{x}_t;\mathbf{z}_{\mathbf{s}_{-N_i}}|\mathbf{z}_{s_i \cup {\mathbf{s}_{N_i}}}) \label{eq:nearpg:wlu:error1}\\
&=c_2(\mathbf{s}_{-i}) + \mathcal{I}(\mathbf{z}_{s_i \cup {\mathbf{s}_{N_i}}};\mathbf{z}_{\mathbf{s}_{-N_i}}) - \mathcal{I}(\mathbf{z}_{s_i \cup {\mathbf{s}_{N_i}}};\mathbf{z}_{\mathbf{s}_{-N_i}}|\mathbf{x}_t) \label{eq:nearpg:wlu:error2}
\end{align}
\end{subequations}
where ${\mathbf{s}_{-N_i}}\triangleq {\mathbf{s}_{1:N}}\setminus\{s_i \cup {\mathbf{s}_{N_i}}\}$ is the set of the chosen sensing locations of the sensors other than sensor $i$ and its neighbors. $c_1(\mathbf{s}_{-i})$ and $c_2(\mathbf{s}_{-i})$ encompass the terms that are constant with respect to the $i$-th sensing agent’s selection, thus they do not affect the preference structure of sensor $i$. 
\begin{proof}
The conditional mutual information of (\ref{eq:pg:wlu:local}) and (\ref{eq:nearpg:wlu:local}) can be expanded using chain rule \cite{Cover1991_Book}
\begin{eqnarray} \label{eq:nearpg:wlu:error:pr} 
  \Delta_{U_i} &= & \mathcal{I}(\mathbf{x}_t;\mathrm{z}_{s_i}|\mathbf{z}_{\mathbf{s}_{N_i}}) - \mathcal{I}(\mathbf{x}_t;\mathrm{z}_{s_i}|\mathbf{z}_{\mathbf{s}_{-i}}) \nonumber \\   
&= &[ \mathcal{I}(\mathbf{x}_t;\mathbf{z}_{\mathbf{s}_{1:N}})-\mathcal{I}(\mathbf{x}_t;\mathbf{z}_{\mathbf{s}_{N_i}})- \mathcal{I}(\mathbf{x}_t;\mathbf{z}_{\mathbf{s}_{-N_i}}|\mathbf{z}_{s_i \cup {\mathbf{s}_{N_i}}})] - [ \mathcal{I}(\mathbf{x}_t;\mathbf{z}_{\mathbf{s}_{1:N}})-\mathcal{I}(\mathbf{x}_t;\mathbf{z}_{\mathbf{s}_{-i}})].
\end{eqnarray}
Cancelling out common terms results in the first equation of (\ref{eq:nearpg:wlu:error}). To obtain the second equation  (\ref{eq:nearpg:wlu:error2}), a preliminary result is given for exchanging of conditioning variables in mutual information \cite{Hoffmann2010_IEEETAC}.
\begin{equation*} 
  \mathcal{I}(\mathrm{x}_a; \mathrm{x}_b|\mathrm{x}_c)  = \mathcal{I}(\mathrm{x}_a;\mathrm{x}_b) - \mathcal{I}(\mathrm{x}_a;\mathrm{x}_c) + \mathcal{I}(\mathrm{x}_a;\mathrm{x}_c|\mathrm{x}_b) 
\end{equation*}
Equation (\ref{eq:nearpg:wlu:error2}) follows by rewriting the third term of (\ref{eq:nearpg:wlu:error:pr}) using the above property. 
\begin{align*}
  \Delta_{U_i} &=\mathcal{I}(\mathbf{x}_t;\mathbf{z}_{\mathbf{s}_{-i}})- \mathcal{I}(\mathbf{x}_t;\mathbf{z}_{\mathbf{s}_{N_i}}) -\mathcal{I}(\mathbf{x}_t;\mathbf{z}_{\mathbf{s}_{-N_i}}) + \mathcal{I}(\mathbf{z}_{s_i \cup {\mathbf{s}_{N_i}}};\mathbf{z}_{\mathbf{s}_{-N_i}}) - \mathcal{I}(\mathbf{z}_{s_i \cup {\mathbf{s}_{N_i}}};\mathbf{z}_{\mathbf{s}_{-N_i}}|\mathbf{x}_t)\\
&=c_2(\mathbf{s}_{-i}) + \mathcal{I}(\mathbf{z}_{s_i}, \mathbf{z}_{\mathbf{s}_{N_i}};\mathbf{z}_{\mathbf{s}_{-N_i}}) - \mathcal{I}(\mathbf{z}_{s_i}, \mathbf{z}_{\mathbf{s}_{N_i}};\mathbf{z}_{\mathbf{s}_{-N_i}}|\mathbf{x}_t).
\end{align*}
\end{proof}
\end{lem}

\begin{rem}
In a target tracking example, the difference (\ref{eq:nearpg:wlu:error2}) between the approximate local utility and the true local utility can be simplified as 
\begin{equation*}
  \Delta_{U_i} = c(\mathbf{s}_{-i}) + \mathcal{I}(\mathbf{z}_{s_i \cup {\mathbf{s}_{N_i}}};\mathbf{z}_{\mathbf{s}_{-N_i}})
\end{equation*} 
Applying the assumption that the measurements are conditionally independent given the target states, the last term in (\ref{eq:nearpg:wlu:error2}) becomes zero, then the error can be represented with mutual information between the sensing selections related with sensing agent $i$ and the others. In this case, the sensing locations that are correlated with sensing agent $i$'s search space are selected as a neighbor set of sensing agent $i$ regardless of the variables of interest. 
\end{rem}

\begin{rem}
In cooperative games, the common terms $c_1(\mathbf{s}_{-i})$ and $c_2(\mathbf{s}_{-i})$ for all the strategies of agent $i$ in (\ref{eq:nearpg:wlu:error}) do not affect its preference structure, because the payoff difference $[U_i(s'_i, \mathbf{s}_{-i})-U_i(s''_i, \mathbf{s}_{-i})]$ quantifies by how much agent $i$ prefers strategy $s'_i$ over strategy $s''_i$ given that the other agents choose the actions $\mathbf{s}_{-i}$ \cite{Candogan2011_MathOR}. Note that a Nash equilibrium is defined in terms of payoff differences, suggesting that games with identical preference structure share the same equilibrium sets. Thus, if the sum of the last two terms in (\ref{eq:nearpg:wlu:error}) is zero, a game with approximate local utility functions will have the same equilibrium sets as a game with a true value of (\ref{eq:pg:wlu:local}).
\end{rem}

\subsection{Geometry-Based Neighbor Selection Algorithm}
Intuitively, the simplest way to determine the neighbor set for each agent is by defining the neighborhood by closeness in geometric distance. In many cases, such as monitoring spatial phenomena as in \cite{Krause2008_JMLR}, geometric closeness of two sensing locations can be translated into correlation between corresponding measurement variables taken at those locations. In this section, we propose a selection algorithm in which neighbors are selected as agents located within some range from each agent. We refer to this selection method a geometry-based neighbor selection. 

One technical assumption made in this section is that the communication network can be represented as a connected undirected graph, $G = (\mathcal{N}, \mathcal{E})$. In this graph, the closeness is defined by the number of hops if multi-hop schemes are used in inter-agent communication. Then, the neighbors of agent $i$ can be represented as $N_i \triangleq \{j:(i,j)\in\mathcal{E}\}$. It is beneficial to limit the neighbors in this way, because the geometry-based selection can obtain both computational and communicational efficiency at the same time. 
First investigate whether or not the localized utility function in (\ref{eq:nearpg:wlu:local}) is aligned with the global potential 
$\phi({s_i},{\mathbf{s}_{-i}})$ in (\ref{eq:pg:wlu:potential}). Using the expression for $\Delta_{U_i}$ in  (\ref{eq:nearpg:wlu:error}), then difference in the utility between two admissible sensing actions $s_i'$ and $s_i''$ can be expressed as:
\begin{align}
\tilde{U}_i(s_i', \mathbf{s}_{-i}) - \tilde{U}_i({s}_i'', {\mathbf{s}_{-i}}) &~= \left[ \phi(s_i', \mathbf{s}_{-i}) -  \phi(s_i'', \mathbf{s}_{-i}) \right] \notag \\
&~~- \underbrace{\left[  \mathcal{I}(\mathbf{x}_t;\mathbf{z}_{\mathbf{s}_{-N_i}}|\mathbf{z}_{s_i' \cup {\mathbf{s}_{N_i}}})
	-  \mathcal{I}(\mathbf{x}_t;\mathbf{z}_{\mathbf{s}_{-N_i}}|\mathbf{z}_{s_i'' \cup {\mathbf{s}_{N_i}}}) \right]}_{\text{difference in potentiality}}
\label{eq:dUi}
\end{align}

From (\ref{eq:dUi}), a game defined by the local utility in (\ref{eq:nearpg:wlu:local}) is not necessarily a potential game with potential in (\ref{eq:pg:wlu:potential}), because the last bracket termed difference in potentiality does not vanish in general\footnote{This statement does not exclude possibility of a potential game with the same local utility function for a different potential function}. Instead, further analysis can be done to identify the condition under which this localized version of the game constitutes a potential game aligned with the objective function of the cooperative sensing.

The last term in (\ref{eq:dUi}) can also be represented as: 
\begin{equation}
\mathcal{I}(\mathbf{x}_t;\mathbf{z}_{\mathbf{s}_{-N_i}}|\mathbf{z}_{s_i \cup {\mathbf{s}_{N_i}}}) = \underbrace{\mathcal{I}(\mathrm{z}_{{s}_{i}}; \mathbf{z}_{\mathbf{s}_{- {N}_i}}  | \mathbf{z}_{\mathbf{s}_{{N}_i}})  - \mathcal{I}(\mathrm{z}_{{s}_{i}}; \mathbf{z}_{\mathbf{s}_{- {N}_i}} |  \mathbf{x}_t, \mathbf{z}_{\mathbf{s}_{{N}_i}})}_{\text{effect of}~ \mathbf{x}_t~\text{on conditional dependency of}~ i~ \text{and}~- {N}_i }
+ c_3(\mathbf{s}_{-i}) \label{eq:delta_Ui}
\end{equation}
by using (\ref{eq:nearpg:wlu:local}).
Noting that the last term in (\ref{eq:delta_Ui}) does not depend on agent $i$'s action, the difference in potentiality represents how significantly the verification variables $\mathbf{x}_t$ changes the conditional dependency between an agent (i.e., $\mathrm{z}_{{s}_i}$) and further agents (i.e., $\mathbf{z}_{\mathbf{s}_{-{N}_i}}$) conditioned on the nearby agents (i.e., $\mathbf{z}_{\mathbf{s}_{{N}_i}}$). One meaningful case is when  $ \mathcal{I}(\mathrm{z}_{{s}_{i}}; \mathbf{z}_{\mathbf{s}_{- {N}_i}}  | \mathbf{z}_{\mathbf{s}_{{N}_i}})  = \mathcal{I}(\mathrm{z}_{{s}_{i}}; \mathbf{z}_{\mathbf{s}_{- {N}_i}} |  \mathbf{x}_t, \mathbf{z}_{\mathbf{s}_{{N}_i}}) = 0$.%
%First, consider the first condition where both of the conditional mutual information terms equal to zero:
\begin{prop} \label{prop:cond_indep}
	If $\mathrm{z}_{{s}_i}$ and $\mathbf{z}_{\mathbf{s}_{-{N}_i}}$ are conditionally independent of each other conditioned on $\mathbf{z}_{\mathbf{s}_{\mathcal{N}_i}}$, for all ${s}_i\in\mathcal{S}_i$ for all $i$,
	$$
	\mathcal{I}(\mathrm{z}_{{s}_{i}}; \mathbf{z}_{\mathbf{s}_{- {N}_i}}  | \mathbf{z}_{\mathbf{s}_{{N}_i}}, \mathbf{y})  = 0
	$$
	for any $\mathbf{y}$. Therefore, the localized utility in (\ref{eq:dUi}) constitutes a potential game with the global potential $\phi$ in (\ref{eq:pg:wlu:potential}).
	\begin{proof}
		Proof is straightforward by noting that
		$
		\mathcal{I}(a;b |c) = 0,
		$ if $a$ and $b$ are conditionally independent each other conditioned on $c$.
	\end{proof}
\end{prop}
In other words, if $\mathrm{z}_{{s}_i}$ is conditionally independent of $\mathbf{z}_{\mathbf{s}_{-{N}_i}}$, then once $\mathbf{z}_{\mathbf{s}_{{N}_i}}$ is known there is no information that can be additionally obtained from $\mathbf{z}_{\mathbf{s}_{- {N}_i}}$ to have better idea about $\mathrm{z}_{{s}_i}$. Furthermore, the statement in Proposition \ref{prop:cond_indep} can be further extended as:
\begin{prop}
	If there exists $K$ such that $\mathrm{z}_{{s}_i}$ are conditionally independent of outside of its $K$-th order neighbors when conditioned on the up to $K$-th order neighbors, $\cup_{k=1}^{K} \mathbf{z}_{\mathbf{s}_{{N}^k_i}}$ where ${N}_k^i $ is $k$-th order neighbor, then the localized utility function
	\begin{equation} %
	\tilde{U}_i^K = \mathcal{I}(\mathbf{x}_t;\mathrm{z}_{{s}_i} | \mathbf{z}_{\mathbf{s}_{{N}_i^1}}, \dots \mathbf{z}_{\mathbf{s}_{{N}_i^K}}) \label{eq:dUiK}
	\end{equation} %
	constitutes a potential game with global potential (\ref{eq:pg:wlu:potential}).
\end{prop}

There are some contexts the conditional independence condition is at least approximately satisfied. Many physical phenomena is associated with some notion of time- and length-scale. If the correlation length-scale of the motion of interest is order of inter-agent distance, there will be no too significant dependency between agents far from each other; then, the conditional independence assumption in the previous propositions can be satisfied. This statement certainly is domain and problem dependent; it is not appropriate to make some general statement of satisfaction of the conditional independence assumption.

\iffalse
\paragraph{Case 2} The second case is more tricky to analyze, because the dependency on the verification variables $V$ also needs to be taken into account. Since there is no physical/geometrical assumption on $V$, the argument on the correlation length-scale does not particularly apply to this case. Instead, when all the random variables are Gaussian and $n_i = 1$ for all $i$, and $|V|=1$, a bit further analysis can be done, because then the mutual information can be written in terms of correlation coefficient between the two random variables.

\begin{Cor} \label{Cor:rho}
	For Gaussian case, for arbitrary index pair $(i,j)$ , if
	$$
	(\rho_{ij} \rho_{jv} \rho_{vi})^2 - \left( \rho_{ij}^2 \rho_{jv}^2 + \rho_{jv}^2 \rho_{vi}^2 + \rho_{vi}^2 \rho_{ij}^2 \right) + 2
	\rho_{ij} \rho_{jv} \rho_{vi} = 0
	$$
	with $\rho_{ij}$ being the correlation coefficient between $Z_{\mathbf{s}_i}$ and $Z_{\mathbf{s}_j}$, the potentiality gap is zero.
\end{Cor}

\begin{Rem}
	Supposed that in Corollary \ref{Cor:rho} $j$ belongs outside of $i$'s neighbor, $\mathbf{s}_j \in \mathcal{S}_{-\mathcal{N}(i)}$.
	One trivial case is where $\rho_{ij} = \rho_{jv} = \rho_{vi} = 0$ or $1$, where conditional independence holds or they are perfectly correlated. There can be non-trivial cases that Corollary \ref{Cor:rho} holds. For example, with $\rho_{ij} = 0.5$, $\rho_{jv} = 0.5$, there exists $\rho_{vi} \in [-1,+1]$.
\end{Rem}

\fi

\subsection{Correlation-Based Neighbor Selection Algorithm}  % Determination of the Neighbor Set
As shown in (\ref{eq:nearpg:wlu:error}), the error incurred by the approximation of the local utility function can be considered in two ways. In the first equation, the error is the mutual information between the target states and the measurement variables which is not in the neighbor set of agent $i$. If after conditioning on the measurement selections of sensing agent $i$ and its neighbors, the measurement variables at $\mathbf{s}_{-N_i}$ have little information about the target variables, the error becomes small enough to approximate the local utility function with sufficient accuracy. That is, the measurement variables at $\mathbf{s}_{-N_i}$ have information about the target states which is already included in the measurements at $s_i$ and $\mathbf{s}_{N_i}$. 

For the second way, the error can be considered as the difference between the prior  mutual information and the posterior mutual information conditioning on the target states. This amounts to the mutual information of the variables at $s_i\cup \mathbf{s}_{N_i}$ and $\mathbf{s}_{-N_i}$ projected onto the subspace generated by the target states \cite{Pinsker1964_Book}. 

\begin{algorithm}[b] \label{alg:neighbor_selection}
  \caption{{\sc Neighbors Selection Algorithm for Weather Forecast Example}($i$, $P_0 = P(\mathbf{z}_{\mathcal{S}_{-i}})$, $P_t = P(\mathbf{z}_{\mathcal{S}_{-i}}|\mathbf{x}_t) $)}
  \begin{algorithmic}[1]
    \STATE $\mathbf{s}_{{N}_i} := \emptyset$
    \STATE $\mathbf{s}_{-{N}_i} := \mathcal{S}_{1:N} \setminus \mathcal{S}_i$
    \FOR{ $j \in \{1,\dots, n\}$}
    \FOR{ $y\in \mathbf{s}_{-{N}_i}$}
    \STATE $e_{y} = \log(\frac{P_0(\mathrm{z}_y)}  {P_t(\mathrm{z}_y)})$ 
    \ENDFOR
    \STATE $ y^* = \arg \max_{y\in \mathbf{s}_{-{N}_i}}e_{y}$
    \STATE $\mathbf{s}_{{N}_i} := \mathbf{s}_{{N}_i} \cup y^*$
    \STATE $\mathbf{s}_{-{N}_i} := \mathbf{s}_{-{N}_i} \setminus y^*$
    \STATE $P_0 =P_0(\mathbf{z}_{\mathbf{s}_{-{N}_i}}) - P_0(\mathbf{z}_{\mathbf{s}_{-{N}_i}},\mathrm{z}_{y^*}) P_0(\mathrm{z}_{y^*}, \mathbf{z}_{\mathbf{s}_{-{N}_i}}) / P_0(\mathrm{z}_{y^*})$
    \STATE $P_t =P_t(\mathbf{z}_{\mathbf{s}_{-{N}_i}}) - P_t(\mathbf{z}_{\mathbf{s}_{-{N}_i}},\mathrm{z}_{y^*}) P_t(\mathrm{z}_{y^*},\mathbf{z}_{\mathbf{s}_{-{N}_i}}) / P_t(\mathrm{z}_{y^*})$
    \ENDFOR
\end{algorithmic}
\label{alg:neighbor_selection}
\end{algorithm}

To make the error sufficiently small, the method to select the neighbor set for each agent should be determined. In most  cases, measurement variables taken at close locations are correlated with each other, and measurements taken at distant locations from sensing agent $i$'s search space have little correlation with agent $i$'s selection. Thus, each sensing agent can approximate its utility function by considering the neighbor set consisting of sensing agents close to each agent. However, for example in sensor planning for weather forecasting, the assumption that closeness in Euclidean distance means correlation between variables is broken. Weather dynamics is highly nonlinear and thus the neighbor set should be chosen in different way from usual cases. For nonlinear target dynamics, such as a weather forecast example, a sequential greedy scheme is proposed. Every sensing agent conducts the greedy scheme to determine its neighbor set. The algorithm simply adds sensing agents in sequence, choosing the next sensor which has maximum mutual information about the target states conditioned on the measurement variables of a sensing agent's search space and its pre-selected neighbors. Using the first error bound in (\ref{eq:nearpg:wlu:error}), the algorithm greedily selects the next sensing agent $j$ that maximizes: 
\begin{equation}
  \mathcal{I}(\mathbf{z}_{s_j};\mathbf{x}_t|\mathbf{z}_{s_i \cup {s_{N_i}}})=\mathcal{H}(\mathbf{z}_{s_j}|\mathbf{z}_{s_i \cup {s_{N_i}}})-\mathcal{H}(\mathbf{z}_{s_j}|\mathbf{z}_{s_i \cup {s_{N_i}}},\mathbf{x}_t).
\end{equation} 
Algorithm \ref{alg:neighbor_selection} shows the greedy neighbor selection algorithm. 

%The error term affecting the preference structure of a game is the difference between the last two terms in (\ref{eq:nearpg:wlu:error}). If the neighbor set for agent $i$ is selected so that the difference is sufficiently small, the game with the approximate local utility is near to the proposed potential game and thus expected to have similar static and dynamic properties to a potential game. 

 If we leave out measurement variables that have little correlation with the selection of sensor $i$, we can approximate the local utility function with a small error. This approximation reduces the burden of computation significantly, making the potential game approach to the sensor network planning problem feasible. However, it cannot be guaranteed that the approximate local utility function satisfies the alignment with the global objective, thus a game with an approximate local utility function may not be a potential game. 
Fortunately, the bound to the Nash equilibrium of the potential game can be obtained from the difference in local utility functions. 

%______________________________________
\subsection{Analysis of Approximate Potential Game}
The approximate utility functions cause a change in the preference structure of the players, and break the alignment condition for a potential game. It follows that the game cannot guarantee the existence of a pure Nash equilibrium which exists in the potential game with the true utility functions. %함수의 근사화로 인해 게임의 선호도 구조가 무너지면서 포텐셜 게임에서 성립하던 alignment 조건이 깨지게 됨. 곧 이것은 포텐셜 게임에서 보장하던 pure NE의 존재성을 입증할 수가 없어짐. 
% pure NE이 존재하지 않지만, approximation error를 충분히 작게 유지한다면 potential game과 가까운 성질이 만족하므로 Candogan 논문 참조 
However, if the approximation error in a utility function stays small, we can expect that a game with the approximate utility functions has similar properties to that of the potential game \cite{Candogan2013_ATEC}. Before presenting the theorem for the equilibria of two games that are close in terms of payoffs, we first introduce the following lemma \cite{Candogan2011_MathOR}. 
 
\begin{lem} 
  \label{lem:nearNE}
   Consider two games $\mathcal{G}=\langle\mathcal{N},\{S_i\}_{i\in\mathcal{N}},\{U_i\}_{i\in\mathcal{N}}\rangle$ and $\tilde{\mathcal{G}}=\langle\mathcal{N},\allowbreak\{\mathcal{S}_i\}_{i\in\mathcal{N}},\allowbreak\{\tilde{U_i}\}_{i\in\mathcal{N}}\rangle$ which differ only in their utility functions. If $|U_i(\mathbf{s})-\tilde{U_i}(\mathbf{s})|\leq\Delta_u$ for every $i\in\mathcal{N}$ and $\mathbf{s}\in\mathcal{S}$, then every $\tilde{\epsilon}$-equilibrium of $\tilde{\mathcal{G}}$ is an $\epsilon$-equilibrium of $\mathcal{G}$ for some $\epsilon \leq 2\Delta_u + \tilde{\epsilon}$ (and vice versa).
\end{lem} 

 In case $\mathcal{G}$ is a potential game, we refer to $\tilde{\mathcal{G}}$ as the approximate potential game of $\mathcal{G}$. Two games that are close to each other in their utilities have similar equilibrium. The lemma shows the extent of how close the $\epsilon$-equilibria of two games are.
Specifically, if $\mathcal{G}$ has a Nash equilibrium, then the other game has an $\epsilon$-equilibrium, such that $\epsilon\leq 2\Delta_u$. Therefore, if we make a game sufficiently close to the existing game with a Nash equilibrium, then an equilibrium property of an approximate game can be identified. %If the approximation error in a utility function stays small, we can guarantee a near equilibrium in the game with the approximate utility functions. 
The following theorem shows that the equilibria of two games can be related in terms of the difference in utilities. 

\begin{thm}
\label{thm:exist:approxeq}
Consider a potential game $\mathcal{G}=\langle \mathcal{N}, \{\mathcal{S}_i\}_{i\in\mathcal{N}}, \{U_i\}_{i\in\mathcal{N}}\rangle$ and its approximate game $\tilde{\mathcal{G}}$ with approximate utility functions $\{\tilde{u}_i\}_{i\in\mathcal{N}}$, i.e, $\tilde{\mathcal{G}}=\langle \mathcal{N}, \{\mathcal{S}_i\}_{i\in\mathcal{N}}, \{\tilde{u}_i\}_{i\in\mathcal{N}}\rangle$. Assume that $|U_i(s_i,\mathbf{s}_{-i})-\tilde{u}_i(s_i, \mathbf{s}_{-i})|\le \Delta_u$ for every $i\in\mathcal{N}$ and $(s_i,\mathbf{s}_{-i})\in\mathcal{S}$. Then every Nash equilibrium of $\mathcal{G}$ is an $\epsilon$-equilibrium of $\tilde{\mathcal{G}}$ for some $\epsilon \le 2\Delta_u$. 
\begin{proof}
  Since a Nash equilibrium is an $\epsilon$-equilibrium with $\epsilon=0$, the result follows from Lemma \ref{lem:nearNE}. 
\end{proof}
\end{thm}
 
This result implies that the Nash equilibria of a potential game with the true utility function are included in the approximate equilibria of the game with approximate utility functions. That is, if we let the set of $\epsilon$-equilibrium of the game $\tilde{\mathcal{G}}$ be $\tilde{\mathcal{X}}_{\epsilon}$ and the set of Nash equilibria of the potential game $\mathcal{G}$ be $\mathcal{X}_0$, then $\mathcal{X}_0 \subset \tilde{\mathcal{X}}_{\epsilon}$. Since the potential game has at least one Nash equilibrium, there exists at least one $\epsilon$-equilibrium in the approximate game $\tilde{\mathcal{G}}$. In a special case, Nash equilibria of the potential game can be maintained in its close potential game with approximate utility functions. If the differences in utility for each player between playing its best response and playing the second-best response at a Nash equilibrium is greater than $2\Delta_u$, then Nash equilibria of the potential game can be included in the set of Nash equilibria of the game. More specifically, if for every $\mathbf{s}^*\in\mathcal{X}_0$
\begin{equation*} 
U_i(s_i^*,\mathbf{s}_{-i}^*)-U_i(s_i,\mathbf{s}_{-i}^*) \ge 2\Delta_u
\end{equation*} 
for every $i\in\mathcal{N}$ and for every $s_i\in\mathcal{S}_i$, $s_i\neq s_i^*$, then every Nash equilibrium of $\mathcal{G}$ is a Nash equilibrium of $\tilde{\mathcal{G}}$. In this case, we can guarantee the existence of a Nash equilibrium in the game with approximate utility functions. That is, $\tilde{\mathcal{X}}_0 \neq \emptyset$ and $\mathcal{X}_0 \subset \tilde{\mathcal{X}}_0$ where $\tilde{\mathcal{X}}_0$ is a set of Nash equilibria of $\tilde{\mathcal{G}}$.

\begin{cor} \label{cor:exist:nasheq}
Consider a potential game  $\mathcal{G}=\langle \mathcal{N}, \{\mathcal{S}_i\}_{i\in\mathcal{N}}, \{U_i\}_{i\in\mathcal{N}}, \phi \rangle$ with global objective function defined as (\ref{eq:pg:wlu:potential}) and local utility function defined as (\ref{eq:pg:wlu:local}) for cooperative sensor network planning and its approximate potential game $\tilde{\mathcal{G}}=\langle\mathcal{N},\allowbreak\{\mathcal{S}_i\}_{i\in\mathcal{N}},\allowbreak\{\tilde{U_i}\}_{i\in\mathcal{N}}\rangle$. Suppose $\Delta_u = \max_{i\in\mathcal{N}, s_i\in\mathcal{S}_i} \mathcal{I}(\mathbf{x}_t;\mathbf{z}_{\mathbf{s}_{-N_i}}|\mathbf{z}_{s_i \cup \mathbf{s}_{N_i}})$. If for every $\mathbf{s}^*\in\mathcal{X}_0$
\begin{equation*} 
\mathcal{H}(\mathrm{z}_{{s_i}}|\mathbf{x}_{t}, \mathbf{z}_{{\mathbf{s}_{-i}}^*})-\mathcal{H}(\mathrm{z}_{{s_i}^*}|\mathbf{x}_t, \mathbf{z}_{{\mathbf{s}_{-i}}^*}) \ge 2\Delta_u
\end{equation*} 
for every $i\in\mathcal{N}$ and for every $s_i\in\mathcal{S}_i$, $s_i\neq s_i^*$, then there exists at least one Nash equilibrium in $\tilde{\mathcal{G}}$ and $\mathcal{X}_0\subset\tilde{\mathcal{X}}_0$.
\begin{proof}
  $\Delta_u=\max_{i\in\mathcal{N}, s_i\in\mathcal{S}_i} \mathcal{I}(\mathbf{x}_t;\mathbf{z}_{\mathbf{s}_{-N_i}}|\mathbf{z}_{s_i \cup \mathbf{s}_{N_i}})$ means that the approximation changes the preference structure of $\mathcal{G}$ with the error less than $\Delta_u$ by Lemma \ref{lem:nearpg:error}. Since 
  \begin{eqnarray*}
U_i(s_i^*, \mathbf{s}_{-i}^*) - U_i(s_i,\mathbf{s}_{-i}^*) & = & \mathcal{I}(\mathrm{z}_{{s_i}^*};\mathbf{x}_{t}|\mathbf{z}_{{\mathbf{s}_{-i}}^*})-\mathcal{I}(\mathrm{z}_{s_i};\mathbf{x}_t|\mathbf{z}_{{\mathbf{s}_{-i}}^*})\\
 & =& \mathcal{H}(\mathrm{z}_{{s_i}}|\mathbf{x}_{t}, \mathbf{z}_{{\mathbf{s}_{-i}}^*})-\mathcal{H}(\mathrm{z}_{{s_i}^*}|\mathbf{x}_t, \mathbf{z}_{{\mathbf{s}_{-i}}^*}), 
 \end{eqnarray*}
there exists at least one Nash equilibrium in $\tilde{\mathcal{G}}$ and $\mathcal{X}_0\subset \tilde{\mathcal{X}}_0$ by the above discussion about the set of Nash equilibrium points. 
\end{proof}
\end{cor}

%\begin{rem}
%	add comments about the neighbor selection algorithm. 
%\end{rem}

%We have demonstrated the existence of equilibrium points in the games with approximate utility %functions, but there remains a problem about efficiency of the resulting equilibrium points. %Therefore, 

% equilibrium point를 정의하는 구간을 넓힘으로써 기존에는 Nash equilibrium이 아니었던 strategy profile까지 평형지점으로 찾는 문제가 생긴다. 
%\begin{rem}
% if the differences in utility functions are greater than the some value, then the Nash equilibrium exists in the approximate potential game and also the Nash equilibrium is same as the original one. 
%\end{rem}
% efficiency에 대해서 일반적으로 말할 수는 없지만, (특별한 경우는 제외하고, submodular 같은) log-linear learning을 사용한 경우에는 

%===============================================================
%\section{Learning Algorithm}
%[TODO: Add some comments about the learning algorithm], and thus the Nash equilibria can be found out with the learning algorithms 

%===============================================================
\section{Numerical Examples} \label{sec:Simulations}
The proposed approximation method is demonstrated on two numerical examples with nonlinear models. In the first example we apply the localized utility function to a sensor selection problem for multi-target tracking in which a particle filter is adopted to estimate the target states. Computation of mutual information for a particle filter grows exponentially with the number of sensing points, thus it is impossible to calculate entropy with full information. The first example shows that the approximation method enables a potential game approach to be applied to a sensor planning problem with a large sensor network. The other example is a simplified weather sensor targeting problem. While the first example shows the performance of the approximation local utility itself, the weather forecast example allows comparison of two neighbor selection algorithms: a geometry-based and a correlation-based method. 

%______________________________________
\subsection{Multi-Target Tracking}
%\subsection{Multi-Target Tracking}
To demonstrate performance of the proposed game-theoretic approach under a nonlinear and non-Gaussian sensor model, a multi-target tracking example is investigated. A similar example using the joint multi target probability density (JMPD) technique introduced in \cite{Kreucher2004_CDC} has been referred.

JMPD is a specialized particle filter method which helps estimate the number of multi-targets and state of each target effectively. JMPD approximates the multi-target state with $N$ particles of $[\mathrm{x}_p^i, \mathrm{w}_i]$, which is expressed as following:
\begin{equation*}
\mathrm{x}_p^i = [ \mathrm{x}_{p,1}, \mathrm{x}_{p,2}, \dots, \mathrm{x}_{p,n_i} ]
\end{equation*}
where $\mathrm{x}_p^i$ is estimating multi-target states, $\mathrm{w}_i$ is a likelihood weight of the particle, $n_i$ is the estimated number of targets and $x_{p,k}$ is a state of the $k$th target. To reallistically design a simulation, each agent is  assumed to carry its own particles. Using the JMPD framework, the probability distribution of the multi-target state (which is equivalent to the verification variable) can be approximated as shown below.
\begin{equation*}
p(\mathrm{x}_t) \approx \sum_{i=1}^{N} \mathrm{w}_i \delta (\mathrm{x}_t - \mathrm{x}_p^i)
\end{equation*}

In this example, a fixed sensor network tries to estimate the number and states of multi-targets moving within a 2400m $\times$ 2400m region with nearly constant speed. In this space, image sensors are uniformly arranged to detect targets. Each sensor gives detection (1) or non-detection (0) as a measurement value according to the range-based detection rate as follows:
\begin{equation*}
P_d(r) = P_{d,0} e^{-r/R_0}
\end{equation*}
where $r$ is the distance between the nearest target and the sensor, and $P_{d,0}$ and $R_0$ are sensor parameters.

The objective of this sensor problem for the multi-target tracking is finding a set of sensing points maximizing the mutual information about the multi-target states at a given situation. Similarly, as in the next weather forecasting example, each agent needs to evaluate the local utility to obtain a cooperative solution. The local utility and its approximate version can be evaluated by using (\ref{eq:pg:wlu:local}) and (\ref{eq:nearpg:wlu:local}), respectively. 
\iffalse
Since sensor observations are conditionally independent given a multi-target state, the local utility can be evaluated as below:
\begin{align*}
U_i(s_i,\mathbf{s}_{-i})&= \mathcal{I}(\mathrm{x}_t; \mathrm{z}_{{s}_i}|\mathrm{z}_{\mathbf{s}_{-i}}) = \mathcal{I}(\mathrm{z}_{\mathbf{s}_i}, \mathrm{z}_{\mathbf{s}_{-i}}; \mathrm{x}_t) \notag \\
&= H(\mathrm{z}_{\mathbf{s}_i}, \mathrm{z}_{\mathbf{s}_{-i}}) - H(\mathrm{z}_{\mathbf{s}_i}, \mathrm{z}_{\mathbf{s}_{-i}}| \mathrm{x}_t)
\end{align*}  
where $\mathrm{z}_{\mathbf{s}}$ is a measurement given by detection results for each sensing points. The approximate local utility can be evaluated as follows.
\begin{align*}
\tilde{U}_i(\mathbf{s}_i,\mathbf{s}_{N_i})&= \mathcal{I}(\mathrm{x}_t; \mathrm{z}_{\mathbf{s}_i}|\mathrm{z}_{N_i}) = \mathcal{I}(\mathrm{z}_{\mathbf{s}_i}, \mathrm{z}_{\mathrm{z}_{N_i}}; \mathrm{x}_t) \notag \\
&= H(\mathrm{z}_{\mathbf{s}_i}, \mathrm{z}_{\mathrm{z}_{N_i}}) - H(\mathrm{z}_{\mathbf{s}_i}, \mathrm{z}_{\mathrm{z}_{N_i}}| \mathrm{x}_t)
\end{align*} 
\fi
Mutual information under a particle filter framework can be computed using the Monte Carlo integration technique below (see \cite{Hoffmann2010_IEEETAC} for more detail derivation). Here, the computation of the integral is equivalent to a finite sum since the possible measurement is only 0 or 1.

\begin{equation}
H(\mathrm{z}_s) \approx -\int_{Z}\Big( \sum_{i=1}^{N} \big( \mathrm{w}_i p(\mathrm{z}_s = \mathrm{z} | \mathrm{x}_t = \mathrm{x}_p^i)\big) \Big) \log \Big(\sum_{i=1}^{N}\big(\mathrm{w}_i p(\mathrm{z}_s = z | \mathrm{x}_t = \mathrm{x}_p^i)\big) \Big) d\mathrm{z} \label{eq:pf_entropy} 
\end{equation}
\begin{equation}
H(\mathrm{z}_s | \mathrm{x}_t) \approx -\int_{Z}\Big( \sum_{i=1}^{N} \big( \mathrm{w}_i p(\mathrm{z}_s = \mathrm{z} | \mathrm{x}_t = \mathrm{x}_p^i) \log p(\mathrm{z}_s = \mathrm{z} | \mathrm{x}_t = \mathrm{x}_p^i) \Big) d\mathrm{z}  \label{eq:pf_cond_entropy} 
\end{equation}
Note the variables in the equation; local utility each agent evaluates cannot coincide with each other if each agent does not share the same particles. Especially in a situation when agents have poor information about the multi-target, it is highly probable that each agent will have different particle sets. Then each agent will estimate different values of the utility even with the same observation result, interrupting cooperation among sensor networks. However, this problem can be solved easily by assuming that agents have the same random generator and share the same random seed so that they can generate exactly the same pseudorandom numbers.

\begin{table}[t]
	\begin{center}%\setlength{\extrarowheight}{4pt}
		\caption{Topology of Multi-Target Tracking Example Cases ($a \times b$: $a$ horizontal grid, $b$ vertical grid)} \label{tab:MTT_setup}
		\vspace*{0.1in}
		\begin{tabular}{c||c|c|c|c|c}
			Case & $N$ & $n_i$ & $\mathcal{S}_{1:N}$& $\mathcal{S}_i$ \\ \hline \hline % & Network \\ \hline \hline
			1  & 6 &2& 3 $\times$ 2 & 2 $\times$ 3  \\ %& 1 $\times$ 6 \\
			2  & 6 &2& 6 $\times$ 1 & 1 $\times$ 6  \\ %& 1 $\times$ 6 \\
		\end{tabular}
	\end{center}
\end{table} 

\begin{table}[b]
	\begin{center}%\setlength{\extrarowheight}{4pt}
		\caption{Objective values for six different strategies} \label{tab:mtt_objvalue}
		\vspace*{0.1in}
		\begin{tabular}{c||c|c|c}
			Strategy &~~~ Case 1~~~ &~~~ Case 2~~~ \\ \hline \hline
			Local greedy  & 0.6299 & 0.7370  \\ %& 1 $\times$ 6 \\
			Sequential greedy  & 0.6386 &0.7524 \\
			JSFP-full w/o inertia  & 0.6415 & 0.7558   \\
			JSFP-full w/ inertia & 0.6415 &0.7543\\
			JSFP-appr 2 hop w/ inertia & 0.6415 &0.7543 \\
			JSFP-appr corr w/ inertia & 0.6415 &0.7543 \\
		\end{tabular}
	\end{center}
\end{table} 

\begin{figure}[t]
	\vspace*{-0.in}
	%\sidecaption[t]
	\centerline{\includegraphics[scale=.4]{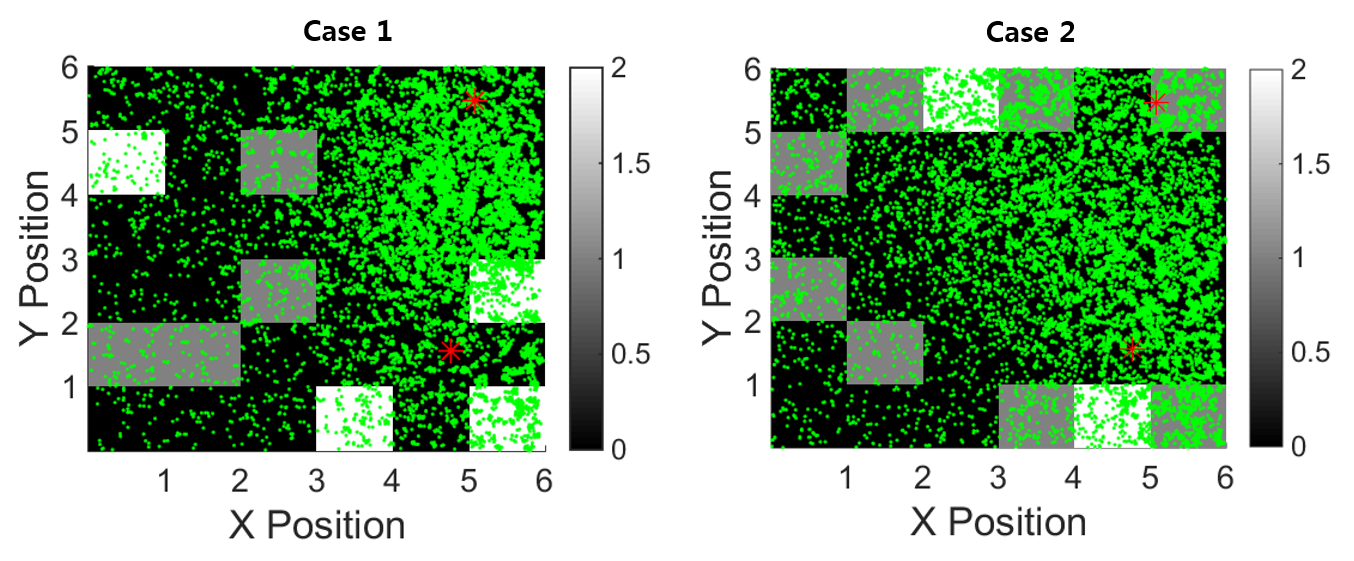}}
	\vspace*{0.1in}
	\caption{Scenario and sensing points determined by the correlation based approximate JSFP in each case. Red stars: true positions of multi-targets. Green dots: particles estimating the state of multi-target}
	\label{fig:mtt_scenario}       % Give a unique label
\end{figure}

The proposed game-theoric method has been tested for two different sensing topologies: 6 sensors in a 3 $\times$ 2 grid and 6 sensors in a 6 $\times$ 1 grid, where each grid corresponds to a 400m $\times$ 400m region over a 2400m $\times$ 2400m space, as described in Table~\ref{tab:MTT_setup}. Each sensor is located in the center of its grid. A sensor can select maximum $n_i$ sensing points, and each sensor can also choose the same sensing points multiple times. None of sensing regions overlap for any two sensors.
\begin{equation} \label{eq:example:nooverlap}
\mathcal{S}_i \cap \mathcal{S}_j =\emptyset, ~~\forall i\neq j 
\end{equation}
As a simulation scenario, a two-target scenario after one observation with each JMPD step is considered. A detail situation is described in Fig. \ref{fig:mtt_scenario}. As shown in the figure, sensors estimate the multi-target states under high uncertainty, thus proper choice of sensing points become important to obtain a large amount of information using the same sensor resources.

\begin{figure}[b]
	\vspace*{-0.in}
	%\sidecaption[t]
	\centerline{\includegraphics[scale=.5]{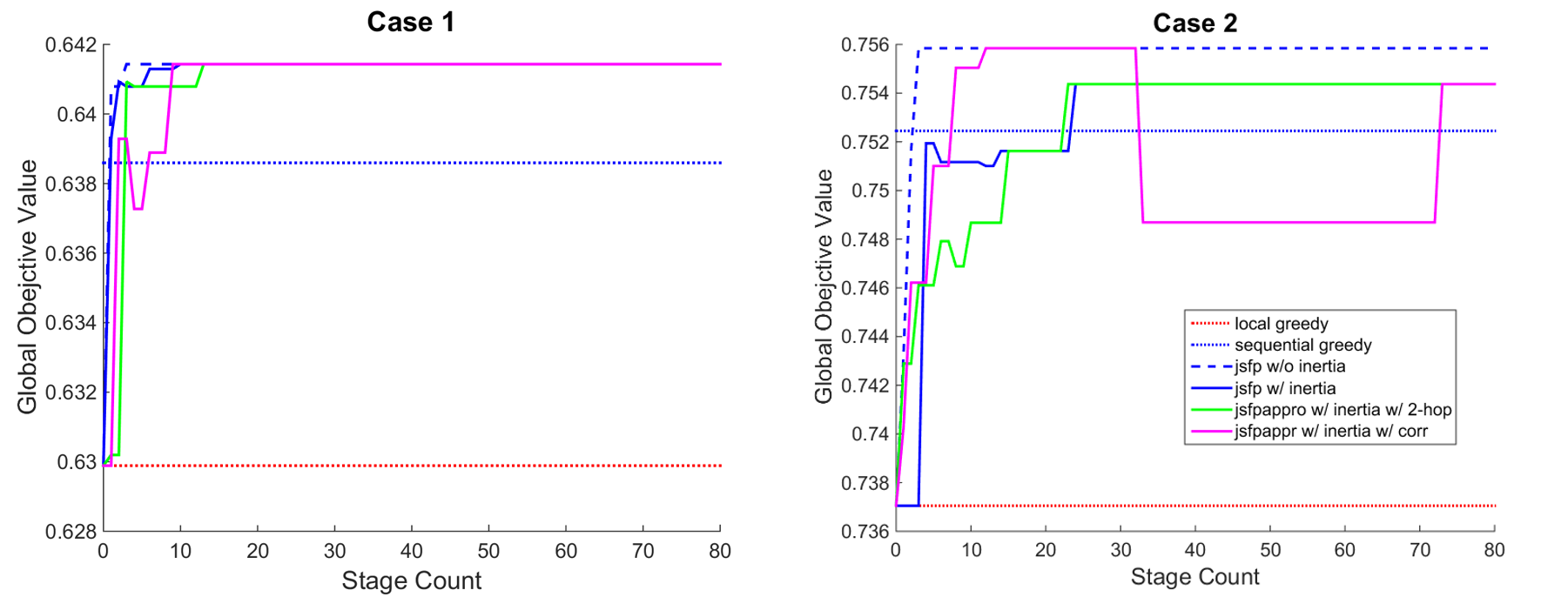}}
	\vspace*{0.1in}
	\caption{Histories of objective values with stage count for two cases}
	\label{fig:mtt_result}       % Give a unique label
\end{figure}

%As in the weather forecasting example, the proposed method is compared with greedy strategies and JSFP algorithm with full information. The global optimum solution is not evaluated since it is hardly obtained in tractable time under particle filter framework. Instead, full information JSFP algorithm has been assumed to give near optimal solution. The resulting objective values for six different strategies are given in Table \ref{tab:mtt_objvalue} and the histories of objective values in the iterative procedure are shown in Fig \ref{fig:mtt_result}. 
The proposed method is compared with greedy strategies and the JSFP algorithm with full information. The global optimum solution is not evaluated since it is hardly obtained in tractable time under the particle filter framework. Instead, full information JSFP algorithm has been assumed to give a near optimal solution. The compared approaches are: 1) greedy strategies in which each agent makes a locally optimal decision using partial information; 2) JSFP learning algorithms in which a local utility is obtained using full information about the other agents' decisions; and 3) JSFP learning rule in which a local utility function is approximated by limiting conditioning variables to the neighbor's decisions. Neighbors are defined in two different ways. One is determined in terms of multi-hop in communication and the other defines neighborhood with correlation in search space. These distributed methods are presented in detail as follows. %in the Appendix. 
\begin{itemize}
	\item Local greedy: Local greedy strategy maximizes the mutual information of its own selection as shown below
	\begin{equation}
	\max \mathcal{I}(\mathrm{z}_{{s}_i};\mathrm{x}_t). \notag
	\end{equation}
	\item Sequential greedy: Each agent selects the sensing location which gives the maximum mutual information conditioned on the preceding agents' decisions. 
	\begin{equation}
	\max \mathcal{I}(\mathrm{z}_{{s}_i};\mathrm{x}_t|\mathbf{z}_{\mathbf{s}_{1:i-1}}) \notag
	\end{equation}
	%\item Iterative greedy: Agents make decisions based on the latest outcome with the same local utility function as (\ref{eq:pg:wlu:local}). The decisions are made iteratively. 
	\item JSFP w/ inertia: Implementation of Algorithm 1 of \cite{Choi2015_IEEETCST} with inertia, i.e., an agent is reluctant to change its action to a better one with some probability (in this example, with probability $\alpha=0.3$ an agent chooses a better action)
	\item JSFP w/o inertia: Implementation of Algorithm 1 of \cite{Choi2015_IEEETCST} without inertia. 
	\item Approximate JSFP with 2-hop neighborhood w/ inertia: Iterative process with local utility functions defined as (\ref{eq:nearpg:wlu:local}). In this strategy, the neighbors are determined in terms of multi-hop in inter-agent communication -- that is, by the geometry-based selection algorithm. 
	\item Approximate JSFP with correlation-based neighborhood w/ inertia: Iterative process with local utility functions defined as (\ref{eq:nearpg:wlu:local}). The neighbors are determined by Algorithm \ref{alg:neighbor_selection} using the correlation structure of the search space. 
\end{itemize}
The resulting objective values for six different strategies are given in Table \ref{tab:mtt_objvalue} and the histories of objective values in the iterative procedure are shown in Fig \ref{fig:mtt_result}. 

Primarily, the JSFP with full information gives the best solution compared to other strategies as expected. In addition, the approximate JSFP gives a better solution than greedy strategies. As seen in case 1, approximate JSFP gives the same  solution as JSFP with full information. However, unlike in the weather forecasting example, the correlation-based approximate JSFP does not always show better performance than the geometry-based approximate JSFP. This is because there is no complex correlation between the sensors, and the correlation is highly related to the Euclidian distance.
Secondly, approximate JSFP algorithms sometimes exhibit an interesting feature in that it may meet the optimal solution during its iteration but does not adopt it, as shown in the correlation-based approximate JSFP of the second example. This is because local utility approximation breaks the potentiality, and agreement among sensors no longer guarantees  cooperation. 
Most importantly, the approximate JSFP algorithm can reduce the computational burden of the JSFP algorithm efficiently while still obtaining a good cooperative solution. Note that calculating the local utility through equations (\ref{eq:pf_entropy}) and (\ref{eq:pf_cond_entropy}) incurs proportional costs to $2^{|s_i \cup \mathbf{s}_i|}$. In other words, computational cost can be reduced exponentially by the number of considering measurement variables $\mathbf{z}_{s_i \cup \mathbf{s}_{N_i}}$ decreased by neighboring approximation. Hence, approximate JSFP can be efficiently used in a particle filter framework with a high order of sensing points which needs increased computation time to evaluate the local utility of JSFP or sequential greedy strategies.

%______________________________________
\subsection{Sensor Targeting for Weather Forecast}\label{subsec:examplemodel}
The proposed game-theoretic method is demonstrated on a sensor targeting example for weather forecast using Lorenz-95 model. The Lorenz-95 model~\cite{Lorenz1998_JAS} is an idealized chaos model that is implemented for the initial verification of numerical weather prediction. In this paper, we consider a  sensor targeting scenario as \cite{Choi2011_IEEETCST}, in which a 2-D extension of the original 1-D Lorenz-95 model was developed and adopted. The 2-D model represents the global weather dynamics of the midlatitude region of the northern hemisphere as follows \cite{Choi2015_IEEETCST}:
\begin{align}
 \dot{y}_{ij} = \left( y_{i+1,j} - y_{i-2,j} \right) y_{i-1,j}
+ \textstyle{\frac{2}{3}} \left( y_{i,j+1} - y_{i,j-2} \right) y_{i,j-1} - y_{ij} + \bar{y},\notag \\ 
(i=1,\ldots,L_{on}, ~j=1,\ldots,L_{at}) \notag
%\label{eq:lorenz}
\end{align}
where $y_{ij}$ denotes a scalar meteorological quantity, such as vorticity or temperature, at the $i$th longitudinal and $j$th latitudinal grid point, each of which corresponds to the state variable at the point. The size of the whole region is $L_{on}=36$ longitudinal and $L_{at}=9$ latitudinal grid points, which are equivalent to 694km $\times$ 694km. 

 The sensor targeting problem for the weather forecast can be rephrased as selecting the optimal sensing locations in the search region at $t_s=0.05$ (equivalent to 6 hours in wall clock) to  reduce the uncertainty in the verification variables. The verification variables correspond to $y$ in the verification region at the verification time $t_v=0.55$ (equivalently to 66 hours). While unattended ground sensors of size 93 are already installed and takes measurements every 6 hours, the decision should be made to choose additional sensing locations for mobile sensing agents, such as UAVs at $t_s$. Using the Ensemble square root filter ~\cite{Whitaker2002_MWR} with the above weather dynamics, the joint probability distribution of the measurement variables $\mathbf{z}_{\mathcal{S}}$ at $t_s$ and the verification variables at $t_v$ can be approximated by a multivariate Gaussian distribution obtained from the samples of the filter (see \cite{Choi2011_IEEETCST} for addional detail of the problem). With the covariance matrix of the measurement variables and the verification variables $P(\mathbf{z}_{\mathcal{S}_{1:N}} \cup \mathrm{x}_t)$, the problem can be treated as a static sensor selection problem. To reduce the computational burden for this sensor selection problem (\ref{eq:globalopt}), a backward scheme ~\cite{Choi2011_IEEETCST} is utilized to calculate the mutual information for the global objective function 
\begin{eqnarray}
  \mathcal{I}(\mathbf{x}_t; \mathbf{z}_{\mathbf{s}}) = \mathcal{I}(\mathbf{z}_{\mathbf{s}}; \mathbf{x}_t) &=& H(\mathbf{z}_{\mathbf{s}}) - H(\mathbf{z}_{\mathbf{s}}|\mathbf{x}_t) \notag \\
  &= &\frac{1}{2} \log (| P(\mathbf{z}_{\mathbf{s}})|) - \frac{1}{2}\log (| P(\mathbf{z}_{\mathbf{s}}|\mathbf{x}_t)|) \notag \\
  & =& \frac{1}{2}\log  \left( \left| P(\mathbf{x}_{\mathbf{s}}) + R_{\mathbf{s}} \right|\right)  - \frac{1}{2}\log \left(\left|P(\mathbf{x}_{\mathbf{s}}|\mathbf{x}_t) + R_{\mathbf{s}} \right|\right). \notag
  %\label{eq:mutinfo:backward}
\end{eqnarray}                                                           
where the measurement equation is given by $\mathrm{z}_s = \mathrm{x}_s + v_s$ with $v_s \sim \mathcal{N}(0, R_s)$ for all $\mathbf{s} \in \mathcal{S}_{1:N}$. For the given covariance matrix $P(\mathbf{x}_{\mathcal{S}_{1:N}} \cup \mathrm{x}_t)$ obtained from the ensemble square root filter, the two covariance matrices $P(\mathbf{x}_{\mathcal{S}_{1:N}}  |\mathrm{x}_t)$ and $P(\mathbf{x}_{\mathcal{S}_{1:N}} )$ are computed prior to the selection process. The unconditioned covariance matrix for the measurement variables $P(\mathbf{x}_{\mathcal{S}_{1:N}} )$ is formed by simply removing the rows and columns corresponding to the verification variables from  $P(\mathbf{x}_{\mathcal{S}_{1:N}} \cup \mathrm{x}_t )$. The conditional covariance matrix $P(\mathbf{x}_{\mathcal{S}_{1:N}}  |\mathrm{x}_t)$ is computed by conditioning $P(\mathbf{x}_{\mathcal{S}_{1:N}} )$ on the verification variables $\mathrm{x}_t$. Once these two covariance matrices are obtained, then the selection process for each sensing agent is equivalent to the selection of the corresponding principal submatrix and calculation of determinants.

 In a potential game, each agent computes the local utility function defined by the conditional mutual information between the measurement selection and the verification variables conditioned on the other agents' action. We obtain the local utility using the backward scheme as the mutual information of the global objective as in (\ref{eq:pg:wlu:localbw}). 
%\begin{align}
% U_i(s_i,\mathbf{s}_{-i})&= \mathcal{I}(\mathbf{x}_t; \mathrm{z}_{s_i}|\mathbf{z}_{\mathbf{s}_{-i}}) = \mathcal{I}(\mathrm{z}_{s_i};\mathbf{x}_t|\mathbf{z}_{\mathbf{s}_{-i}}) \notag \\
%&= H(\mathrm{z}_{s_i}|\mathbf{z}_{\mathbf{s}_{-i}}) - H(\mathrm{z}_{s_i}|\mathbf{x}_t,\mathbf{z}_{\mathbf{s}_{-i}}) \notag \\
%  &= \frac{1}{2} \log \left(\left| P(\mathrm{z}_{s_i}|\mathbf{z}_{\mathbf{s}_{-i}})\right|\right) - \frac{1}{2}\log \left(\left| P(\mathrm{z}_{s_i}|\mathbf{x}_t,\mathbf{z}_{\mathbf{s}_{-i}})\right|\right)
%  \label{eq:condmutinfo:backward}
%\end{align}  
We should calculate the two matrices $P(\mathbf{z}_{\mathcal{S}_i}|\mathbf{z}_{\mathbf{s}_{-i}})$ and $ P(\mathbf{z}_{\mathcal{S}_i}|\mathbf{x}_t,\mathbf{z}_{\mathbf{s}_{-i}})$ over the search space  of agent $i$ before comparing the preference of the agent. 
For the obtained covariance matrices $P(\mathbf{x}_{\mathcal{S}_{1:N}}  |\mathrm{x}_t)$ and $P(\mathbf{x}_{\mathcal{S}_{1:N}} )$ from the backward scheme the two conditional covariance  matrices  $P(\mathbf{z}_{\mathcal{S}_i}|\mathbf{z}_{\mathbf{s}_{-i}})$ and $ P(\mathbf{z}_{\mathcal{S}_i}|\mathrm{x}_t,\mathbf{z}_{\mathbf{s}_{-i}})$ are computed by conditioning on the other agents' sensing selections respectively. If the number of sensing points each agent selects is one, then the covariance matrix for one sensing point becomes a scalar which is a corresponding diagonal elements in the matrix.

  The approximate local utility is computed in the same way as computing the local utility (\ref{eq:pg:wlu:localbw}) with the exception of the conditioning variables. The conditioning variables are replaced with the neighboring agents instead of all the other agents' selections. 
\begin{align}
 \tilde{U}_i(\mathbf{s}_i,\mathbf{s}_{N_i})&=  \mathcal{I}(\mathrm{z}_{\mathbf{s}_i};\mathrm{x}_t|\mathrm{z}_{\mathbf{s}_{N_i}}) \notag \\
&= H(\mathrm{z}_{\mathbf{s}_i}|\mathrm{z}_{\mathbf{s}_{N_i}}) - H(\mathrm{z}_{\mathbf{s}_i}|\mathrm{x}_t,\mathrm{z}_{\mathbf{s}_{N_i}}) \notag \\
  &= \frac{1}{2} \log \left(\left| P(\mathrm{z}_{\mathbf{s}_i}|\mathrm{z}_{\mathbf{s}_{N_i}})\right|\right) - \frac{1}{2}\log \left(\left| P(\mathrm{z}_{\mathbf{s}_i}|\mathrm{x}_t,\mathrm{z}_{\mathbf{s}_{N_i}})\right|\right)
  \label{eq:apprcondmutinfo:backward}
\end{align}   

%\paragraph{Comparative Results}\label{subsec:exampleresult}

The proposed game-theoretic method using an approximation of the local utility has been tested for three different sensing topologies; nine sensors in a 3 $\times$ 2 format in two different search spaces, and fifteen sensors in a 2 $\times$ 3 format in a larger region than the first and second cases, as described in Table~\ref{tab:sim_setup}. An oceanic region of size $12 \times 9$ (in longitude $\times$ latitude) is considered as a potential search region, among which the whole search space $\mathcal{S}_{1:N}$  is chosen and each agent is assigned its own sensing region $\mathcal{S}_i$ separated from the other agents as shown in (\ref{eq:example:nooverlap}) of the multi-target tracking example. 
%\begin{equation}
%\mathcal{S}_i \cap \mathcal{S}_j =\emptyset, ~~\forall i\neq j \notag
%\end{equation}
The number of sensing locations for each agent is set to be one for all the cases, as in \cite{Choi2015_IEEETCST}, because the global optimal solution cannot be obtained in tractable time. Several approaches, both centralized and distributed, are compared with the proposed method. The centralized approach is an exact method that searches the entire candidate space $\mathcal{S}$, by which we can obtain the global optimal solution. As in the multi-target tracking example, the distributed approaches are greedy strategies and JSFP algorithm with full information.

\begin{table}[t]
\begin{center}%\setlength{\extrarowheight}{4pt}
\caption{Topology of Example Cases ($a \times b$: $a$ grids in Longitude, $b$ grids in Latitude)} \label{tab:sim_setup}
\vspace*{0.1in}
\begin{tabular}{c||c|c|c|c|c}
    Case & $N$ & $n_i$ & $\mathcal{S}_{1:N}$& $\mathcal{S}_i$ \\ \hline \hline % & Network \\ \hline \hline
1  & 9 &1& 9 $\times$ 6 & 3 $\times$ 2  \\ %& 1 $\times$ 6 \\
2  & 9 &1& 9 $\times$ 6 & 3 $\times$ 2  \\ %& 1 $\times$ 6 \\
3  & 15 &1& 10 $\times$ 9 & 2 $\times$ 3  \\% & 2 $\times$ 3 \\
\end{tabular}
\end{center}
\end{table} 

\begin{figure}[t]
\vspace*{-0.in}
%\sidecaption[t]
\centerline{\includegraphics[scale=.5]{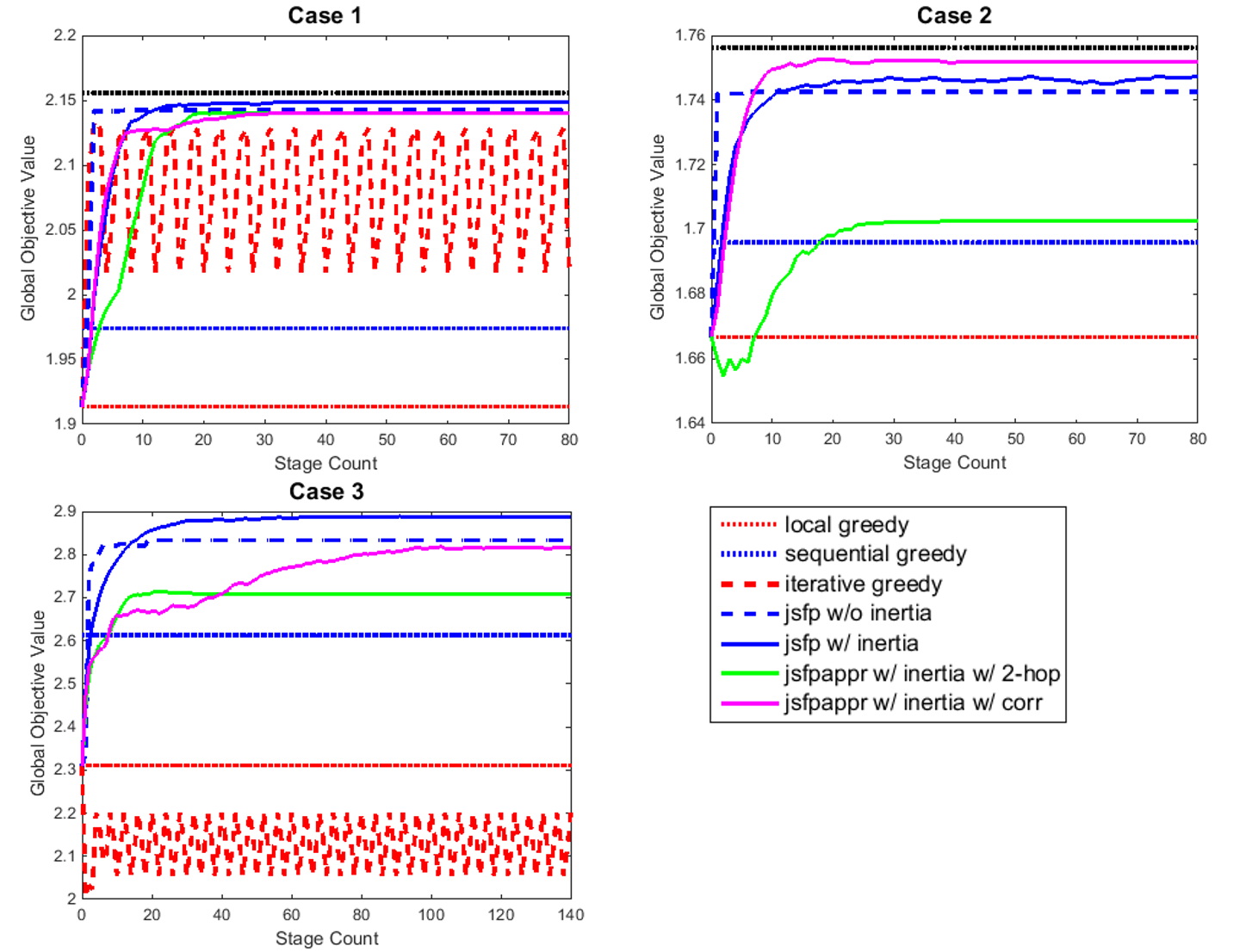}}
\vspace*{0.1in}
\caption{Histories of objective values with stage count for three cases}
\label{fig:historyobjvalue}       % Give a unique label
\end{figure}

The resulting objective values for the seven different strategies are given in Table~\ref{tab:objvalue}, and the histories of objective values in the iterative procedure are shown in Fig.~\ref{fig:historyobjvalue}. The results for iterative algorithms with inertia are obtained from Monte-Carlo simulation and represent average objective values. Case 3 is different from the other two cases in that a larger sensor network is considered, and the global optimal solution cannot be obtained in tractable time. However, from the examples of small sensor networks we consider that the optimal solution of the third case may be close to the JSFP with full information. Thus, we can consider the objective value for the JSFP with full information as an upper bound for the sensor networks with a large number of sensors.
% The sequential greedy strategy can be considered as a baseline for comparing the performance of different strategies, since it guarantees the worst-case performance in polynomial time, even though the guarantee is applied to the problems in which the objective functions satisfy some conditions. That is, the sequential greedy algorithm specifis a lower bound for performance. Thus, the proposed method for finding out an approximate solution should give an objective value higher than sequential greedy algorithm's solutions. 

\begin{table}[t]
\begin{center}%\setlength{\extrarowheight}{4pt}
\caption{Objective values for seven different strategies} \label{tab:objvalue}
\vspace*{0.1in}
\begin{tabular}{c||c|c|c}
    Strategy &~~~ Case 1~~~ &~~~ Case 2~~~ &~~~ Case 3~~~  \\ \hline \hline
Global optimal  & 1.6222 & 1.7563 & N/A  \\ %& 1 $\times$ 6 \\
Local greedy  & 1.3941 & 1.6668 & 2.8238  \\ %& 1 $\times$ 6 \\
Sequential greedy  & 1.4301 &1.6959 & 3.0218 \\
JSFP-full w/o inertia  & 1.6222 &1.7427& 3.2319   \\
JSFP-full w/ inertia & 1.6162 &1.7479&3.2236 \\
JSFP-appr 2 hop w/ inertia & 1.4393 &1.7026&2.9476 \\
 JSFP-appr corr w/ inertia & 1.5928 &1.7519&3.1662 \\
\end{tabular}
\end{center}
\end{table} 

First, note that the JSFP with full information of other agents' actions gives a better solution than the sequential greedy algorithm, and the solution is the same as or close to the optimal solution in \cite{Choi2015_IEEETCST}. The sequential greedy strategy can be considered a baseline for comparing the performance of different strategies, since it guarantees the worst-case performance in polynomial time, even though the guarantee is applied to problems in which the objective functions satisfy some conditions. The JSFP with approximate local utility functions also presents a better performance than the sequential greedy strategy except for approximate JSFP with 2-hop neighborhood of the third example. The approximate local utility function based on the correlation always gives a better solution than ones depending on the actions of the neighbors selected by geometry-based method. In cases 1 and 2, the objective value for the approximate JSFP with a correlation-based neighborhood is close to the JSFP with full information. The important thing to note here is that the number of conditioning variables used for computing the utility functions is half of the JSFP with full information. As mentioned in Sec.~\ref{sec:comp_time_analysis}, the computation time for the conditional mutual information increases proportional to the cubic of the conditioning variables. Therefore, the computation time for the approximate local utility function is reduced by a factor of approximately 8. 

% TODO: 영진이것과 비교해서 어떤 때는 좋고, 어떤 때는 나쁘다. 그래서 correlation이 더 좋아진다. 

%===============================================================
\section{Conclusion}
An approximation method for computing the local utility function has been presented when a sensor network planning problem is formulated as a potential game. A local utility function of each agent that depends on the neighboring agents' decisions is presented, and a neighbor selection algorithm is proposed to keep the error induced by the approximation small. Two sensor targeting examples for weather forecasting and multi-target tracking demonstrated that potential game formulation with the approximation local utility function gives good performance close to a potential game with full information. % and results in a better solution than previous work, in which the approximation local utility function depending on the neighboring agents' actions was proposed but the neighboring agents are specified in terms of physical distance. % depends on the actions of the neighboring agents located within a certain distance. 

\bibliographystyle{IEEEtran}
\bibliography{Reference}

\begin{thebibliography}{10}
\providecommand{\url}[1]{#1}
\csname url@rmstyle\endcsname
\providecommand{\newblock}{\relax}
\providecommand{\bibinfo}[2]{#2}
\providecommand\BIBentrySTDinterwordspacing{\spaceskip=0pt\relax}
\providecommand\BIBentryALTinterwordstretchfactor{4}
\providecommand\BIBentryALTinterwordspacing{\spaceskip=\fontdimen2\font plus
\BIBentryALTinterwordstretchfactor\fontdimen3\font minus
  \fontdimen4\font\relax}
\providecommand\BIBforeignlanguage[2]{{%
\expandafter\ifx\csname l@#1\endcsname\relax
\typeout{** WARNING: IEEEtran.bst: No hyphenation pattern has been}%
\typeout{** loaded for the language `#1'. Using the pattern for}%
\typeout{** the default language instead.}%
\else
\language=\csname l@#1\endcsname
\fi
#2}}

\bibitem{Choi2015_IEEETCST}
H.-L. Choi and S.-J. Lee, ``A potential-game approach for
  information-maximizing cooperative planning of sensor networks,''
  \emph{{IEEE} Transactions on Control Sytems Technology}, vol.~23, no.~6, pp.
  2326--2335, November 2015.

\bibitem{Gopalakrishnan2011_ACM}
R.~Gopalakrishnan, J.~R. Marden, and A.~Wierman, ``An architectural view of
  game theoretic control,'' \emph{ACM SIGMETRICS Performance Evaluation
  Review}, vol.~38, no.~3, pp. 31--36, 2011.

\bibitem{Hoffmann2010_IEEETAC}
G.~M. Hoffmann and C.~J. Tomlin, ``Mobile sensor network control using mutual
  information methods and particle filters,'' \emph{{IEEE} Transactions on
  Automatic Control}, vol.~55, no.~1, pp. 32--47, January 2010.

\bibitem{Krause2008_JMLR}
A.~Krause, A.~Singh, and C.~Guestrin, ``Near-optimal sensor placements in
  gaussian processes: Theory, efficient algorithms and empirical studies,''
  \emph{Journal of Machine Learning Research}, vol.~9, no.~2, pp. 235--284,
  June 2008.

\bibitem{Nguyen2016_IEEETCST}
L.~Nguyen, S.~Kodagoda, R.~Ranasinghe, and G.~Dissanayake, ``Information-driven
  adaptive sampling strategy for mobile robotic wireless sensor network,''
  \emph{IEEE Transactions on Control Systems Technology}, vol.~24, no.~1, pp.
  372--379, 2016.

\bibitem{Grocholsky2002_Thesis}
B.~Grocholsky, ``Information-theoretic control of multiple sensor platforms,''
  Ph.D. dissertation, University of Sydney, 2002.

\bibitem{Boyd2011_FTML}
S.~Boyd, N.~Parikh, E.~Chu, B.~Peleato, and J.~Eckstein, ``Distributed
  optimization and statistical learning via the alternating direction method of
  multipliers,'' \emph{Foundations and Trends{\textregistered} in Machine
  Learning}, vol.~3, no.~1, pp. 1--122, 2011.

\bibitem{Marden2009_IEEETAC}
J.~R. Marden, G.~Arslan, and J.~S. Shamma, ``Joint strategy fictitious play
  with inertia for potential games,'' \emph{{IEEE} Transactions on Automatic
  Control}, vol.~54, no.~2, pp. 208--220, February 2009.

\bibitem{Marden2007_ICAAMS}
------, ``Regret based dynamics: convergence in weakly acyclic games,'' in
  \emph{Proceedings of the 6th international joint conference on Autonomous
  agents and multiagent systems}.\hskip 1em plus 0.5em minus 0.4em\relax ACM,
  2007, p.~42.

\bibitem{Candogan2011_MathOR}
O.~Candogan, I.~Menache, A.~Ozdaglar, and P.~Parrilo, ``Flows and
  decompositions of games: Harmonic and potential games,'' \emph{Mathematics of
  Operations Research}, vol.~36, no.~3, pp. 474--503, 2011.

\bibitem{Choi2013_ACC}
H.-L. Choi, ``A potential game approach for distributed cooperative sensing for
  maximum mutual information,'' in \emph{Proc. American Control Conference},
  Washington DC, USA, 2013.

\bibitem{Cover1991_Book}
T.~M. Cover and J.~A. Thomas, \emph{Elements of Information Theory}.\hskip 1em
  plus 0.5em minus 0.4em\relax Wiley-Interscience, 1991.

\bibitem{Marden2009_IEEETSMC}
J.~R. Marden, G.~Arslan, and J.~S. Shamma, ``Cooperative control and potential
  games,'' \emph{{IEEE} Transactions on Systems, Man, and Cybernetics, Part
  B(Cybernetics)}, vol.~39, no.~6, pp. 1393--1407, Dec. 2009.

\bibitem{Choi2016_DARS}
H.-L. Choi, K.-S. Kim, L.~B. Johnson, and J.~P. How, ``Potential game-theoretic
  analysis of a market-based decentralized task allocation algorithm,'' in
  \emph{Distributed Autonomous Robotic Systems}.\hskip 1em plus 0.5em minus
  0.4em\relax Springer, 2016, pp. 207--220.

\bibitem{Fudenberg1991_Book}
D.~Fudenberg and J.~Tirole, \emph{Game Theory}.\hskip 1em plus 0.5em minus
  0.4em\relax Cambridge, MA: MIT Press, 1991.

\bibitem{Monderer1996_Games}
D.~Monderer and L.~Shapley, ``Potential games,'' \emph{Games and Econom.
  Behav.}, vol.~14, no.~1, pp. 124--143, May 1996.

\bibitem{Fudenberg1998_Book}
D.~Fudenberg and D.~K. Levine, \emph{The theory of learning in games}.\hskip
  1em plus 0.5em minus 0.4em\relax MIT Press, 1998.

\bibitem{Choi2011_IEEETCST}
H.-L. Choi and J.~P. How, ``Efficient targeting of sensor networks for
  large-scale systems,'' \emph{{IEEE} Transactions on Control Sytems
  Technology}, vol.~19, no.~6, pp. 1569--1677, November 2011.

\bibitem{Borowski2013_IEEECDC}
H.~Borowski, J.~R. Marden, and E.~W. Frew, ``Fast convergence in semi-anonymous
  potential games,'' in \emph{52nd IEEE Conference on Decision and
  Control}.\hskip 1em plus 0.5em minus 0.4em\relax IEEE, 2013, pp. 2418--2423.

\bibitem{Andersen2002_LectureNotes}
B.~Andersen, J.~Gunnels, F.~Gustavson, and J.~Was'niewski, ``A recursive
  formulation of the inversin of symmetric positive definite matrices in packed
  storage data format,'' \emph{Lecture Notes in Computer Science}, vol. 2367,
  pp. 287--296, 2002.

\bibitem{Pinsker1964_Book}
M.~Pinsker, \emph{Information and Information Stability of Random Variables and
  Processes}.\hskip 1em plus 0.5em minus 0.4em\relax Oxford: Holden-Day, 1964.

\bibitem{Candogan2013_ATEC}
O.~Candogan, A.~Ozdaglar, and P.~A. Parrilo, ``Near-potential games: Geometry
  and dynamics,'' \emph{ACM Transactions on Economics and Computation}, vol.~1,
  no.~2, p.~11, 2013.

\bibitem{Kreucher2004_CDC}
C.~Kreucher, A.~O. Hero, K.~Kastella, and D.~Chang, ``Efficient methods of
  non-myopic sensor management for multitarget tracking,'' in \emph{Decision
  and Control, 2004. CDC. 43rd IEEE Conference on}, vol.~1.\hskip 1em plus
  0.5em minus 0.4em\relax IEEE, 2004, pp. 722--727.

\bibitem{Lorenz1998_JAS}
E.~Lorenz and K.~Emanuel, ``Optimal sites for supplementary weather
  observations: simulation with a small model,'' \emph{Journal of the
  Atmospheric Sciences}, vol.~55, no.~3, pp. 399--414, 1998.

\bibitem{Whitaker2002_MWR}
J.~Whitaker and H.~Hamill, ``Ensemble data assimilation without perturbed
  observations,'' \emph{Monthly Weather Review}, vol. 130, no.~7, pp.
  1913--1924, 2002.

\end{thebibliography}
\end{document}